\documentclass[epjc3,twocolumn,nopacs]{svjour3}
\pdfoutput=1
\usepackage[utf8]{inputenc}
\usepackage{amsmath,amsfonts,amssymb,slashed,graphicx,hyperref,color,cite,xspace,orcidlink}
\usepackage{multirow,siunitx,enumitem,arydshln,nicefrac}
\hypersetup{colorlinks=true,linkcolor=blue,citecolor=magenta,filecolor=magenta,urlcolor=cyan}
\journalname{Eur. Phys. J. C}
\graphicspath{{figures/}}
\newcommand{\be}{\begin{equation}}
\newcommand{\ee}{\end{equation}}
\def\bsp#1\esp{\begin{split}#1\end{split}}

\newcommand{\delphes}{{\sc Del\-phes~3}\xspace}
\newcommand{\fastjet}{{\sc FastJet}\xspace}
\newcommand{\hepmc}{{\sc HepMC}\xspace}
\newcommand{\madanalysis}{{\sc Mad\-A\-na\-ly\-sis~5}\xspace}
\newcommand{\madgraph}{{\sc MG5\_aMC}\xspace}
\newcommand{\hepdata}{{\sc HEPData}\xspace}
\newcommand{\sarah}{{\sc Sarah}\xspace}
\newcommand{\python}{{\sc Python}\xspace}
\newcommand{\pythia}{{\sc Pythia~8}\xspace}
\newcommand{\sampleanalyzer}{{\sc SampleAnalyzer}\xspace}
\newcommand{\checkmate}{{\sc Check\-MA\-TE~2}\xspace}

\def\invfb{\ensuremath{{\rm fb}{}^{-1}}\xspace}
\def\ie{{\it i.e.}}
\def\eg{{\it e.g.}}
\def\etc{{\it etc.}}

\begin{document}
\title{Recasting LHC searches for long-lived particles with \madanalysis}
\author{
  Jack Y. Araz\thanksref{e1,addr1}\orcidlink{0000-0001-8721-8042}
  \and
  Benjamin Fuks\thanksref{e2,addr2}\orcidlink{0000-0002-0041-0566}
  \and
  Mark D. Goodsell\thanksref{e3,addr2}\orcidlink{0000-0002-6000-9467}
  \and
  Manuel~Utsch\thanksref{e4,addr2}
}

\thankstext{e1}{E-mail: {\color{cyan}jack.araz@durham.ac.uk}}
\thankstext{e2}{E-mail: {\color{cyan}fuks@lpthe.jussieu.fr}}
\thankstext{e3}{E-mail: {\color{cyan}goodsell@lpthe.jussieu.fr}}
\thankstext{e4}{E-mail: {\color{cyan}utsch@lpthe.jussieu.fr}}

\institute{
    Institute for Particle Physics Phenomenology, Durham University, South Road, Durham, DH1 3LE, UK\label{addr1}
   \and\
    Laboratoire de Physique Th\'eorique et Hautes Energies (LPTHE),
    UMR 7589, Sorbonne Universit\'e et CNRS, 4 place Jussieu,
    75252 Paris Cedex 05, France\label{addr2}
}

\date{Received: date / Accepted: date}

\maketitle

\begin{abstract}
We present an extension of the simplified fast detector simulator of \madanalysis\ -- the \emph{SFS framework} -- with methods making it suitable for the treatment of long-lived particles of any kind. This allows users to make use of intuitive \python commands and straightforward C++ methods to introduce detector effects relevant for long-lived particles, and to implement selection cuts and plots related to their properties. In particular, the impact of the magnetic field inside a typical high-energy physics detector on the trajectories of any charged object can now be easily simulated. As an illustration of the capabilities of this new development, we implement three existing LHC analyses dedicated to long-lived objects, namely a CMS run~2 search for displaced leptons in the $e\mu$ channel (CMS-EXO-16-022), the full run~2 CMS search for disappearing track signatures (CMS-EXO-19-010), and the partial run~2 ATLAS search for displaced vertices featuring a pair of oppositely-charged leptons (ATLAS-SUSY-2017-04). We document the careful validation of all \madanalysis SFS implementations of these analyses, which are publicly available as entries in the \madanalysis Public Analysis Database and its associated dataverse.
\end{abstract}

\maketitle

\section{Introduction}\label{sec:intro}
In operation for more than a decade, the Large Hadron Collider (LHC) has already collected a rich amount of data. Whereas this has allowed for a clear establishment of the existence of a particle consistent with the Standard Model (SM) Higgs boson, the experimental evidence of its \emph{potential} and even whether it is a fundamental scalar is still lacking. There is also no compelling sign for physics beyond the SM (BSM), and even if there are currently several suggestive hints, these are not obviously connected to electroweak symmetry breaking. Before the LHC operation, it was hoped that not only would the Higgs boson be found, but that other particles which would be discovered and play a role in protecting the electroweak scale from the hierarchy problem. The lack of evidence therefore suggests that new phenomena may be more complicated to grasp than was hoped for by phenomenologists and experimentalists alike, and hiding in unexpected or underexploited corners of the parameter space. In this context, the high-energy physics community has shown a growing theoretical and experimental interest in long-lived particles (LLPs)~\cite{Alimena:2019zri}.

Naive dimensional analysis suggests that a particle should have a decay length $c\tau$ inversely proportional to its mass $M$:
\be\bsp
c \tau \sim& \SI{10}{fm} \left(\frac{\mathrm{GeV}}{M} \right). 
             \esp\ee
There are naturally three mechanisms whereby this expectation can be violated: decays mainly occurring at loop order (for example, the neutral pion decaying to two photons); the particle having small couplings; and there being little phase space for the decay (\eg\ if three-body decays dominate and/or the particle spectrum is compressed). The smallness of the naive decay length has led to the large majority of new physics searches at the LHC focusing on promptly decaying new particles. However, the SM contains many long-lived particles: most famously the muon with a proper decay length of \SI{660}{m}; but also of huge importance are the $B$ mesons, which are the heaviest LLPs in the SM, with masses of about 5~GeV. These latter decay mainly via three-body processes which are severely suppressed by weak couplings and the Cabibbo–Kobayashi–Maskawa (CKM) matrix, and so have $c\tau \sim \SI{0.5}{mm}$. This demonstrates just how naive the above expectation is, and indeed particles featuring a long lifetime due to loop, coupling or phase space suppression are also found in a large variety of BSM setups. Heavy LLPs therefore make striking signals with low backgrounds, and offer a path to new discoveries in run 3 of the LHC.

A number of search strategies already exist, targeting different kinds of LLP candidates. If we focus on the case of the LHC run~2 and the usage of the general-purpose ATLAS and CMS experiments, LLPs are searched for through displaced vertices in the inner detector or in the muon spectrometer~\cite{ATLAS:2018tup,ATLAS:2019kpx,ATLAS:2019fwx,ATLAS:2019jcm,ATLAS:2020wjh,ATLAS:2020xyo,ATLAS:2021jig,CMS:2018tuo,CMS:2021tkn,CMS:2021juv}, displaced jets originating from the calorimeters~\cite{ATLAS:2018niw,ATLAS:2019qrr,CMS:2018qxv,CMS:2019qjk,CMS:2020iwv}, displaced lepton jets~\cite{ATLAS:2019tkk,CMS:2019dqq}, emerging jets~\cite{CMS:2018bvr}, non-pointing and delayed photons~\cite{CMS:2019zxa}, disappearing tracks~\cite{ATLAS:2017oal,ATLAS:2021ttq,CMS:2018rea,CMS:2020atg}, stop\-ped particles~\cite{ATLAS:2021mdj,CMS:2017kku} and heavy stable charged particles~\cite{ATLAS:2018imb,ATLAS:2018lob,ATLAS:2019gqq,ATLAS:2019wkg}. Reusing these searches and applying them to different models -- not just those considered by the experimental analyses -- via a user-friendly computer program would be highly valuable. For that reason, allowing this possibility in the \madanalysis package~\cite{Conte:2012fm,Conte:2014zja,Conte:2018vmg} has recently become a priority in the development of the program.

The first steps in this direction date from 2018, when \madanalysis version 1.6 was released. This consisted of the first version of the code that allowed for handling of displaced leptons~\cite{Fuks:2018yku}, and it was the first time that one of the public recasting codes was made suitable for the recasting of an LHC analysis involving displaced objects. In practice, the recasting module of the code had been updated in a twofold manner. First, it was linked to a modified version of \delphes~\cite{deFavereau:2013fsa} for the simulation of the detector response. These modifications, which were merged shortly later with the official version of the \delphes detector simulator package, allowed the user to go beyond the default behaviour in which all particles are assumed to decay within the tracker volume of the detector. Second, information about the impact parameter and the position from which a lepton originated was incorporated into the internal data format of the code. As a proof of concept, the CMS-EXO-16-022 analysis~\cite{CMS:2016isf} was implemented in \madanalysis~\cite{C6B9V3_2021}. This implementation was validated in a satisfactory manner, so that for most signal regions \madanalysis reproduced the expected signal for the model given within 20\%. However, for long decay lengths ($\ge 100$ cm), the difference between the \madanalysis result and the expected signal was greater than $100\%$.

Subsequently, due to the intense experimental and theoretical interest in this area, additional public codes became available. Firstly, a collaborative collection of independent stand-alone codes, called the {\tt LLP Re\-cast\-ing Repository}, was established on the {\sc GitHub} platform~\cite{llprecasting}. It has recently been augmented by two sophisticated disappearing track searches \cite{Belyaev:2020wok,Goodsell:2021iwc}. Secondly, an LLP version of \checkmate\cite{Desai:2021jsa}, which includes five (classes of) LLP searches, and an updated version of {\sc SModelS}~\cite{Alguero:2021dig}, which includes eight LLP sear\-ches, were released.

Last year, the Simplified Fast Detector Simmulation (SFS) framework of \madanalysis was introduced. It consists of a new simplified fast detector simulator relying on efficiency and smearing functions~\cite{Araz:2020lnp}. Additional commands were supplemented to the \python command line interface of \madanalysis, so that users could design and tune a detector simulator directly from a set of intuitive commands, and the C++ core of the program was accordingly modified to deal with such a detector simulator. Proof-of-concept examples were presented both for simple phenomenological analyses and in the context of the reinterpretation of the results of existing LHC analyses in refs.~\cite{Araz:2020lnp,Fuks:2021zbm,Araz:2020stn}, and additional works demonstrated its utility in more in-depth studies~\cite{Araz:2020zyh,Banerjee:2021oxc,Fuks:2021xje,Chakraborty:2021tdo,Deandrea:2021vje,Goodsell:2021iwc}.

In this work, we further extend the SFS framework and the core of \madanalysis by equipping them with additional methods suitable for the treatment of long-lived objects of any kind. Section~\ref{sec:ma5} is dedicated to a brief review of the \madanalysis framework and its fast detector simulator module, the SFS, as well as a detailed description of all new developments suitable for LLP studies. In section~\ref{sec:CMS_LLP_analysis}, we make use of these developments to implement, in the \madanalysis SFS framework, the CMS-EXO-16-022~\cite{CMS:2016isf} search for displaced leptons in LHC run~2 data. Such a search was already implemented earlier, using \delphes as a detector simulator, and consists thus of an excellent benchmark to evaluate and compare the performance of the SFS simulator to recast LHC searches for long-lived particles. In section~\ref{sec_ATLAS_LLP_analysis} we describe the recast of a new run 2 analysis~\cite{ATLAS:2019fwx} from the ATLAS Collaboration searching for displaced vertices consisting of two (visible) leptons. Section~\ref{sec_CMS_DT} describes the implementation in \madanalysis SFS of the full run 2 CMS disappearing track search \cite{CMS:2018rea,CMS:2020atg}. We summarise and conclude in section~\ref{sec:conclusion}. All our implementations are publicly available on the \madanalysis dataverse~\cite{SVILPJ_2021,31JVGJ_2021,P82DKS_2021} and its Public Analysis Database~\cite{Dumont:2014tja}.\footnote{See the webpage \url{http://madanalysis.irmp.ucl.ac.be/wiki/PublicAnalysisDatabase}.}

\section{\madanalysis as a tool for detector simulation and LHC recasting}\label{sec:ma5}
\subsection{Generalities}\label{sec:ma5general} \setcounter{equation}{0}

\madanalysis~\cite{Conte:2012fm,Conte:2014zja,Conte:2018vmg} is a general framework dedicated to different aspects of BSM phenomenology. First, it offers various means for users to design a new analysis targeting any given collider signal. To this aim, the code is equipped with a user-friendly \python-based command-line interface and a developer-friendly C++ core. Any user can then either implement their analysis through predefined functionalities available from the interpreter of the program, relying on the {\it normal mode} of running, or implement their needs directly in C++ in the {\it expert mode} of the code. Second, \madanalysis allows users to automatically reinterpret the results of various LHC searches for new physics to estimate the sensitivity of the LHC to an arbitrary BSM signal. The list of available searches and details about the validation of their implementation can be obtained through the Public Analysis Database (PAD) of \madanalysis. Moreover, recent developments allow for automated projections at higher luminosities and the inclusion of theoretical uncertainties~\cite{Araz:2019otb}, and leave the choice to simulate the response of the detector with either the external \delphes package~\cite{deFavereau:2013fsa} or the built-in SFS framework~\cite{Araz:2020lnp}.

The SFS module of \madanalysis has been designed to emulate the response of a detector through a joint usage of the \fastjet software~\cite{Cacciari:2011ma} and a set of user-defined smearing and efficiency functions. In this way, the code can effectively mimic typical detector effects such as the impact of resolution degradations and object reconstruction and identification. In practice, \madanalysis post-processes the set of objects resulting from the application of a jet algorithm to the hadron clusters in a Monte Carlo event. Reconstruction efficiencies, tagging efficiencies and the corresponding mistagging rates are implemented as transfer functions inputted in the \madanalysis command-line interface. In addition, the latter also allows users to provide the standard deviation associated with the potential Gaussian smearing of the four-momentum of a given object, such smearing representing any reconstruction degradations. All those transfer functions can be general piecewise functions and depend on several variables such as the components of the four-momentum of the object considered.

\subsection{Features and new developments of the SFS}\label{sec:ma5general}
\subsubsection{The SFS framework in a nutshell}
The detector effects described in the previous subsection are implemented through dedicated \madanalysis comma\-nds that are extensively described in the SFS manual~\cite{Araz:2020lnp}. We therefore only provide below a brief description of the extant features, emphasising instead the new developments that are documented for the first time in this work. Instructions related to how to run the code are available from ref.~\cite{Conte:2018vmg}.

In order to turn on the SFS module, users have to start \madanalysis in the reconstructed mode ({\tt ./bin/ma5 -R}) and switch on reconstruction by means of \fastjet. This is achieved by typing in\footnote{This assumes that \fastjet is available on the system. If this is not the case, it can be installed locally by typing in the \madanalysis command-line interface {\tt install fastjet}.}
\begin{verbatim} set main.fastsim.package = fastjet\end{verbatim}
By default, objects will be reconstructed by making use of the anti-$k_T$ algorithm~\cite{Cacciari:2008gp}, with a radius parameter set to $R=0.4$. The jet algorithm considered and the associated properties can be modified through the commands
\begin{verbatim} set main.fastsim.algorithm = <algo>
 set main.fastsim.<property> = <value>\end{verbatim}
where \verb|<algo>| represents the keyword associated with the jet algorithm adopted, and {\tt <property>} generically denotes any of its properties. For instance,
\begin{verbatim} set main.fastsim.radius = 1.0\end{verbatim}
would set the jet radius parameter of the anti-$k_T$ algorithm to $R=1$. At this stage, hadron-level events have to be provided through files encoded either in the standard {\sc HepMC} format~\cite{Dobbs:2001ck} or in the deprecated {\sc StdHep} format that is still supported in \madanalysis. Event files can be imported one by one or simultaneously (through wildcards in the path to the samples) by typing in the command-line interface of \madanalysis
\begin{verbatim} import <path-to-hepmc-files> as <set>\end{verbatim}
This command can be repeated as many times as necessary when multiple event files have to be imported. According to the above syntax, all event samples are collected into a single dataset defined by the label {\tt <set>}, although multiple datasets (defined through different labels) can obviously be used as well. We refer to refs.~\cite{Conte:2018vmg,Araz:2020lnp} for detailed information about all available options. 

In addition, users must indicate to the code the set of objects that are invisible and leave no track or hit in a detector, and the set of objects that participate in the hadronic activity in the event. This is achieved by typing in
\begin{verbatim} define invisible = invisible <pdg-code>
 define hadronic = hadronic <pdg-code>\end{verbatim}
where, in the above examples, one adds a new state defined through its Particle Data Group (PDG) identifier~\cite{ParticleDataGroup:2020ssz} ({\tt pdg-code}) to the {\tt invisible} and {\tt hadronic} containers. These containers include by default the relevant Standard Model particles and hadrons, as well as the supersymmetric lightest neutralino and gravitino for what concerns {\tt invisible}.

In order to include in this \fastjet-based reconstruction process various detector effects, the SFS comes with three main commands,
\begin{verbatim} define reco_efficiency <obj> <fnc> [<dom>]
 define smearer <obj> with <obs> <fnc> [<dom>]
 define tagger <obj> as <reco> <fnc> [<dom>]\end{verbatim}
The first of these commands indicates to the code that the reconstructed object {\tt <obj>} can only be reconstruc\-ted with a given efficiency. The latter is provided as a function {\tt <fnc>} that can be either a constant or depend on the properties of the object (its transverse momentum $p_T$, its pseudorapidity $\eta$, \etc). If the optional argument {\tt <dom>} is not provided, this function applies to the entire kinematic regime. Otherwise, the efficiency is a piecewise function that is defined by issuing the above command several times, each occurrence coming with a different function {\tt <fnc>} and a different domain of definition {\tt <dom>} (specified through inequalities).

The second command instructs \madanalysis that the property {\tt <obs>} of the object {\tt <obj>} has to be smea\-red in a Gaussian way, the resolution being provided as a function {\tt <fnc>} of the properties of the object. Such a function can again be given as a piecewise function by issuing the {\tt define smearer} command several times, for different functions {\tt <fnc>} and domains of definition {\tt <dom>}. The last command above enables users to provide the efficiency to tag (or mistag) an object {\tt <obj>} as a (possibly different) object {\tt <reco>}. The efficiencies are given through a function {\tt <fnc>} of the object properties, the command being issued several times with the optional argument {\tt <dom>} when piecewise functions are in order.

\subsubsection{New developments}
Relative to its initial release 1.5 years ago, the current public version of the SFS simulator (shipped with \madanalysis version 1.9) has been augmented with respect to three aspects.

First, new options are available for the treatment of photon, lepton and track isolation. Users can specify the size $\Delta R$ of an isolation cone that will be associated with these objects by typing, in the command line interface,
\begin{verbatim} set main.fastsim.<prt>_isocone_radius = \
    <value>\end{verbatim}
Here, {\tt <prt>} should be replaced by {\tt track}, {\tt electron}, {\tt muon} or {\tt photon}, and \verb|<value>| can be either a floating-point number or a set of comma-separated floating point numbers. In the former case, it is associated with the $\Delta R$ value of the isolation cone, whereas in the latter case isolation information for several values of the isolation cone radius are computed on runtime. The SFS module then takes care of computing the sum of the transverse momentum of hadrons, photons and charged leptons located inside a cone of radius $\Delta R$ centred on the object considered ($\sum p_T$), as well as the amount of transverse energy $\sum E_T$ inside the cone. At the level of the analysis implementation, all isolation cone objects are available, together with the corresponding $\sum p_T$ and $\sum E_T$ information. We refer to \ref{app:propexpert} for practical details. These new features are used in practice in the analysis of section \ref{sec_CMS_DT} where validation was performed by comparing with implementations using Monte Carlo truth information also available to the user. 

Secondly, the SFS module now includes an energy scaling option. Such a capability is known to be useful in numerous analyses, in particular when jet properties are in order. Users can instruct the code to rescale the energy of any given object by typing, in the \madanalysis interpreter,
\begin{verbatim} define energy_scaling <obj> <fnc> [<dom>]\end{verbatim}
Similarly to the other SFS commands, this syntax indicates to the code that the energy of the object {\tt <obj>} has to be rescaled. This object can be an electron ({\tt e}), a muon ({\tt mu}), a photon ({\tt a}), a hadronic tau ({\tt ta}) or a jet ({\tt j}). Once again, the keyword {\tt <fnc>} is a proxy for the functional form of the scaling function, and  {\tt <dom>} consists of an optional attribute relevant for piecewise functions. In the specific case of jet energy scaling (JES), users can rely on the equivalent syntax,
\begin{verbatim}
 define jes <fnc> [<dom>]
\end{verbatim}
This feature is not used by the analyses presented in this paper, however, but is intended for future applications.

Monte Carlo event records, usually stored in the \hepmc format~\cite{Dobbs:2001ck}, include the exact positions of the vertices related to the (parton-level and hadron-level) cascade decays originating from the hard-scattering process,\footnote{When using \madgraph the user must ensure that this information is included by changing the {\tt time\_of\_flight} variable in {\tt run\_card.dat}; by default vertex positions are discarded.} as well as the momenta of all intermediate- and final-state objects. This information is however computed by neglecting the impact of the presence of a magnetic field in the detector volume, ignoring the fact that charged particle tracks are typically bent under the influence of such a magnetic field. Subsequently, this can indirectly affect the trajectories of any electrically neutral particle that would emerge from a decay chain involving a charged state. While such effects are negligible in many cases, the computational effort to simulate them exactly is reasonable compared with the remaining workload in the analysis tool chain.

We have therefore equipped the SFS framework with a module handling particle propagation in the presence of a magnetic field. It can be turned on by typing, in the \madanalysis command line interpreter,
\begin{verbatim}
 set main.fastsim.particle_propagator = on
\end{verbatim}
the default value being {\tt off}. In the default case, all objects propagate along a straight line. When the propagation module is turned on, the direction of the momentum of any char\-ged particle or object and its decay products is modified. The SFS module assumes that all objects produced in the event are subjected to a homogeneous magnetic field parallel to the $z$-axis, with a magnitude that can be fixed through
\begin{verbatim}
  set main.fastsim.magnetic_field = <value>
\end{verbatim}
In this last command, {\tt value} stands for any positive floating-point number. On runtime, the SFS module evaluates the trajectory of any electrically-charged object as derived from the Lo\-rentz force originating from the field. The code next calculates the coordinates of the point of closest approach, which corresponds to the point of the trajectory at which the distance to the $z$-axis is smallest, and it finally extracts the values of the transverse and longitudinal impact parameters $d_0$ and $d_z$. The formulas internally used are derived in detail in \ref{app:propagation}.

Each coordinate of the point of closest approach is accessible, in the \madanalysis command line interpreter, through the {\tt XD, YD, ZD} symbols. The transverse and longitudinal impact parameters $d_0$ and $d_z$ are associated with the {\tt D0} and {\tt DZ} symbols, and their approximated variants in the case of straight-line propagation $\tilde{d}_0$ and $\tilde{d}_z$ are mapped with the {\tt D0APPROX} and {\tt DZAPPROX} symbols. All those symbols can in addition be manipulated as for any other observable implemented in \madanalysis. For instance, the scalar difference between the transverse impact parameter of two objects can be obtained by prefixing {\tt sd} to the name of the observable (\ie\ through the {\tt sdD0} symbol in the example considered). We refer to the manual for more information on observable combinations~\cite{Conte:2012fm}.

The set of commands available in the normal mode of \madanalysis now support the above variables. It is hence possible to plot (via the command {\tt plot} of the \madanalysis interpreter) these observables. Users can also use them in the implementation of selection cuts (through the complementary {\tt select} and {\tt reject} commands of the \madanalysis interpreter), and of course in the SFS framework to define LLP reconstruction and tagging efficiencies. In the expert mode of the code, access to this information is provided as described in \ref{app:propexpert}.

\subsection{Technicalities on particle propagation in the SFS}
When turned on, the particle propagation module of \madanalysis assumes the existence of a non-zero magnetic field along the $z$-axis whose magnitude is fixed by the user. It evaluates the impact of this magnetic field on the trajectories of all intermediate- and final-state objects stored in a hadron-level event record. The implemented algorithm begins with the final-state particles originating from the hard process, and then follows the structure of all subsequent radiation and decays through the mother-to-daughter relations encoded in the event record. The primary interaction vertex is taken as located at the origin of the reference frame, and the hard process is considered to occur at $t=0$.

For any specific object, the propagation algorithm makes use of information on the object's creation position ${\bf x}_{\rm creation}$ and time $t_{\rm creation}$ (from the decay of the mother particle or the hard-scattering process), as well as on the time of its decay $t_{\rm decay}$. The propagation time $\Delta t_{\rm prop}$ is thus given by
\begin{equation}
  c \Delta t_{\rm prop} = c t_{\rm decay} - c t_{\rm creation}.
\end{equation}
The position ${\bf x}_{\rm decay}$ at which the object's decay occurs is then estimated from eq.~\eqref{eq:trajectory}, the momenta of the daughter particles are rotated according to eq.~\eqref{eq:p} and the impact parameters $d_0$ and $d_z$ are computed from eq.~\eqref{eq:impactparameters}. The code additionally calculates approximate values for the impact parameters, as would be derived when assuming that the object's trajectory is a straight line, and information on the momentum rotation angle is consistently passed to all subsequent objects emerging from the object's (cascade) decay. The propagation of the decay products originating from ${\rm x}_{\rm decay}$ is next iteratively treated.

It may happen that some objects have two mother particles (like when hadronisation processes are in order). The evaluation of the position ${\bf x}_{\rm creation}$ and time $t_{\rm creation}$ is then potentially ambiguous. However, such a case is always associated with zero lifetimes and is thus irrelevant for the problem considered in this work.

\begin{figure*}\centering
  \includegraphics[width=.49\textwidth]{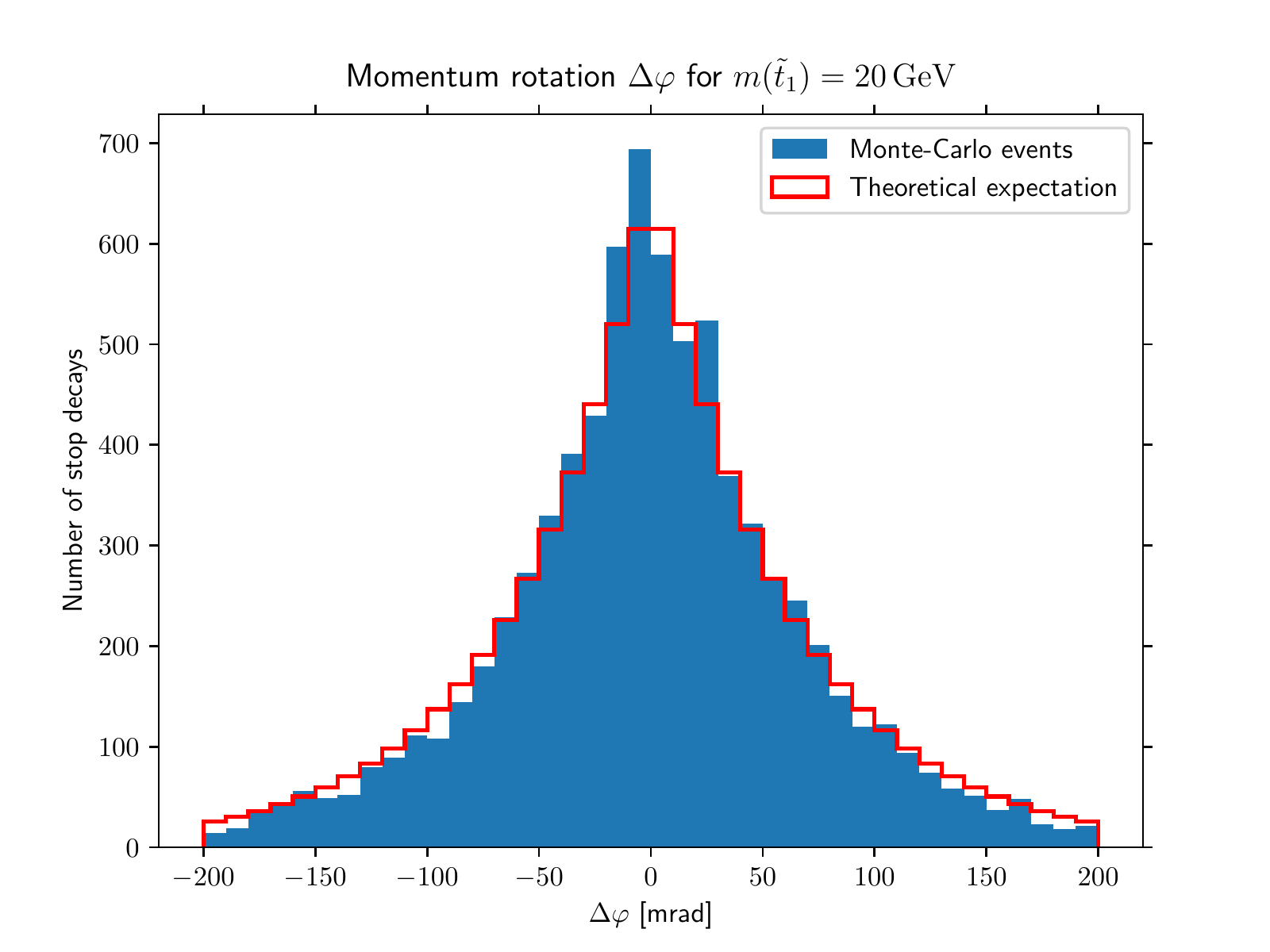}
  \includegraphics[width=.49\textwidth]{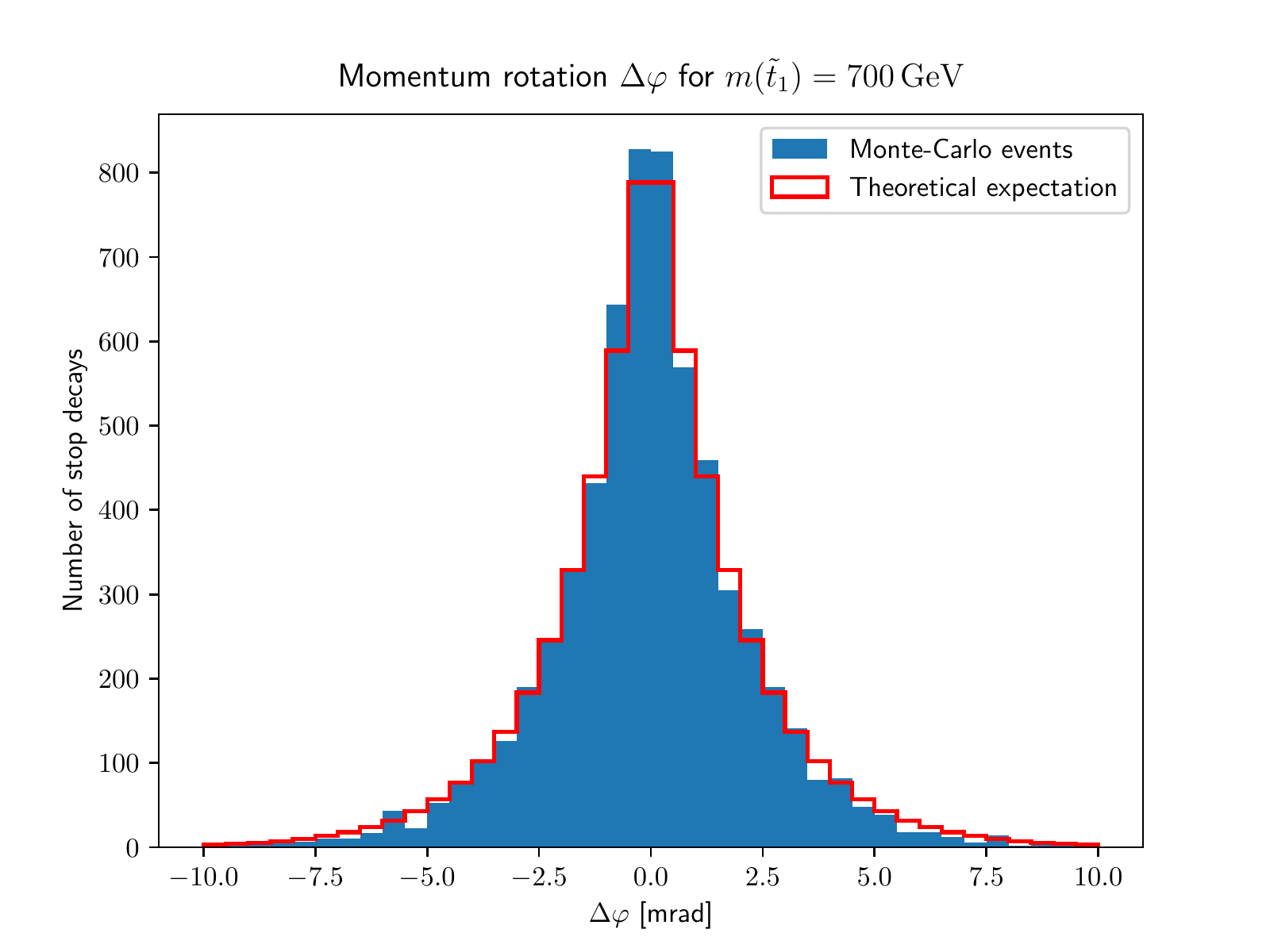}\\
  \includegraphics[width=.49\textwidth]{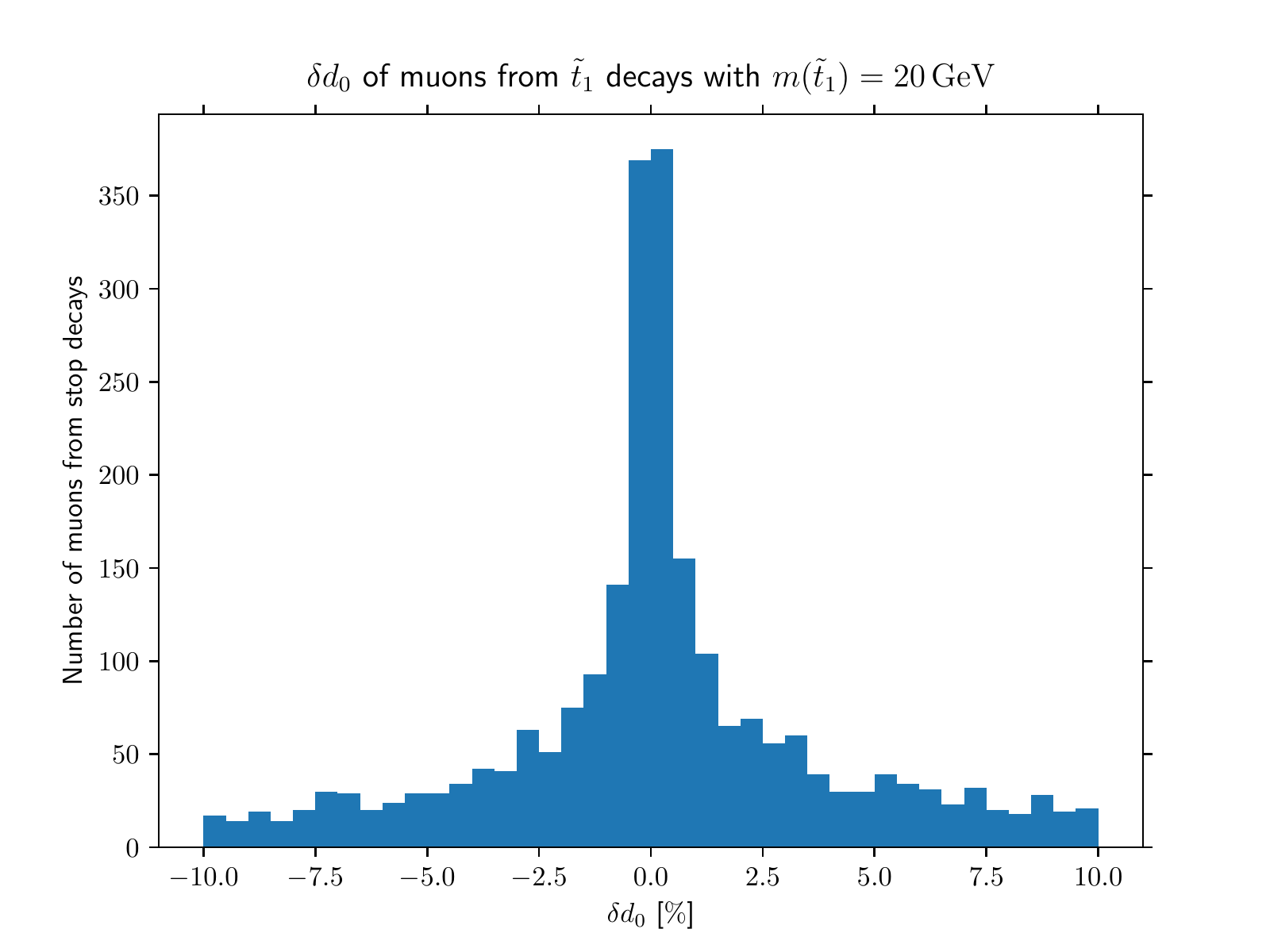}
  \includegraphics[width=.49\textwidth]{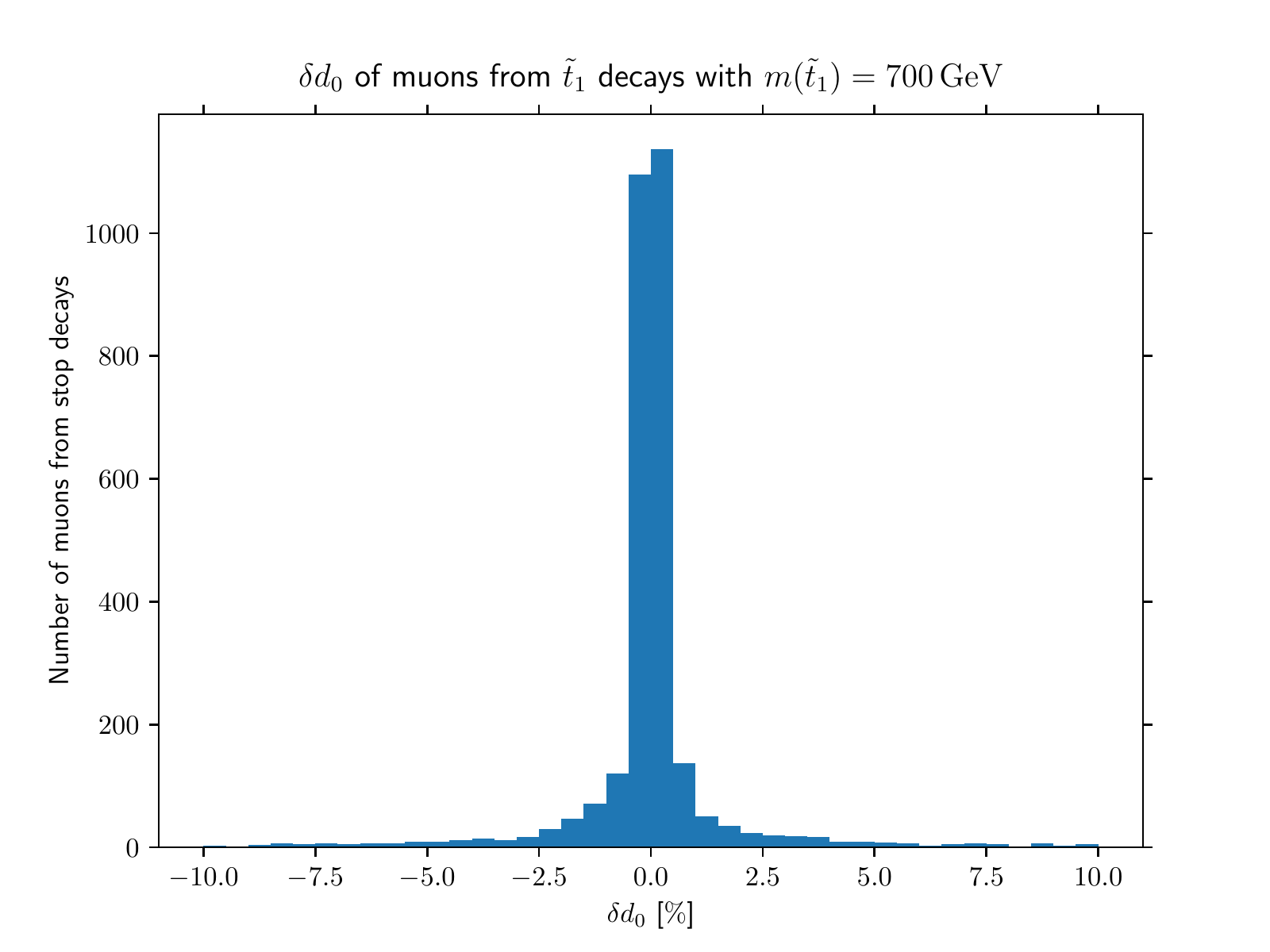}
  \caption{Distribution in the angle $\Delta\varphi$ (top row) associated with the rotation to which the momentum of the $R$-hadrons originating from the process~\eqref{eq:validprocess} is subjected, when they propagate in a magnetic field of 4~T. The associated deviation $\delta d_0$ on the transverse impact parameter of the final-state muons is additionally shown (bottom row). We consider a scenario where the stop mean decay length is set to 1~m, and where the stop mass is fixed to 20~GeV (left) or 100~GeV (right).\label{fig:ma5_valid}
  }
\end{figure*}

\subsection{Validation of the SFS propagation module}
The validation of the implementation of object propagation in the SFS framework has been carried out through a study of stop pair production and decay in $R$-parity violating supersymmetry~\cite{Barbier:2004ez,Graham:2012th}. We consider the process
\be pp\to \tilde t\tilde t^* \to b \ell^+ \ \bar b\ell^{\prime -}, \label{eq:validprocess}\ee
where each of the pair-produced top squarks can decay into an electron or a muon with equal branching ratio. The stop is taken to be long-lived, so that it hadronises into an $R$-hadron of about the same mass which then flies some distance in the detector before decaying. This leads to the presence of displaced leptons in the final state of each event, as well as to the possible appearance of electrically charged and unstable intermediate particles (\ie\ the produced $R$-hadrons), which are sensitive to an external magnetic field when they propagate. In particular, the momentum of the produced $R$-hadrons will rotate by an angle $\Delta\varphi$ during their propagation. From the formulas presented in \ref{app:propagation}, we can analytically extract the $\Delta\varphi$ distribution for a given stop mass $m_{\tilde t}$ (which is a good proxy for the $R$-hadron masses) and decay time $t_{\rm decay}$, 
\be \Delta\varphi = \frac{q_R B t_{\rm decay}}{m_{\tilde t}}.\ee
Here, $q_R$ stands for the $R$-hadron electric charge. The expected $\Delta\varphi$ distribution originating from the probability distribution of the decay time can then be compared with predictions obtained with \madanalysis and its SFS module.

For this purpose, we make use of \pythia~\cite{Sjostrand:2014zea} to generate 50,000 signal events for two scenarios respectively featuring a stop mass of 20 and 700~GeV, and in both of which the stop mean decay length has been fixed to 1~m.\footnote{In the context of the validation of our implementation, we are only interested in the physics associated with given scenarios, and not whether these scenarios are excluded by current data.} In the lightest configuration, we expect more enhanced magnetic field effects than in the heaviest setup, due to the larger related cyclotron frequency.

The hadronisation is also performed in \pythia, which includes a treatment of the formation of $R$-ha\-drons. In the following, since we are interested in the impact of the magnetic field, we do not consider any special effects of the $R$-hadron interaction with the detector, and treat them as propagating freely in the magnetic field until they decay. Indeed, in the SFS approach these would need to be handled by efficiency functions, prossibly on an analysis by analysis basis (see \eg~ref.~\cite{ATLAS:2019gqq}).

In the top row of figure~\ref{fig:ma5_valid}, we show predictions for the $\Delta\varphi$ distribution obtained when the magnetic field is set to 4~T. We compare results obtained from our analytical calculations (solid red line) to those returned by \madanalysis when particle propagation is turned on (blue histograms). Excellent agreement is found, and we can observe that the spectrum is much broader in the case of the lightest scenario. This therefore confirms that
the propagator module works as expected. 

We assess the effect of the magnetic field on the final-state lepton by examining the shift that is induced on the impact parameter spectrum. For illustrative purposes, we compute the relative difference between the transverse impact parameter $d_0$ as obtained when particle propagation is turned on, and its $\tilde d_0$ variant assuming straight-line propagation,
\begin{equation}
  \delta d_0=\frac{d_0-\tilde d_0}{\tilde d_0}.
\end{equation}
We present results for muons in the bottom row of figure~\ref{fig:ma5_valid}, the electron counterparts being found similar. As already mentioned, the mass of the long-lived $R$-hadrons (or equivalently the mass of the top squark) plays an important role in the shape of the distribution. The impact is indeed stronger for lighter $R$-hadrons, as the objects that originate from their decay automatically have a trajectory that is more bent. Although our findings depict possibly important effects, they do not necessarily have important consequences on a typical LLP LHC analysis once the new physics goals in terms of a mass range are defined. This however should be investigated on a case-by-case basis.

The new particle propagator module is used by the recast analyses that we present in the following sections. However, the impact of the magnetic field is not especially significant. There may be other cases in the future where it is important, for example searches for slowly moving particles. Of more importance in the following is that the code now includes methods for correctly treating displaced vertices, track objects, computing $d_0$, \etc

\section{Displaced leptons in the {\boldmath $e\mu$} channel (CMS-EXO-16-022)\label{sec:CMS_LLP_analysis}}
\setcounter{equation}{0} \subsection{Generalities}
The experimental analysis presented in ref.~\cite{CMS:2016isf} is a search for long-lived particles where each event produces two displaced leptons, precisely one electron and one muon. It was carried out by the CMS Collaboration in 2015, \ie\ at the beginning of the run-2 data-taking period of the LHC at $\sqrt{s}=\SI{13}{TeV}$ and with a data sample corresponding to an integrated luminosity of $\SI{2.6}{fb^{-1}}$. This search superseded the \SI{8}{TeV} analysis described in ref.~\cite{Khachatryan:2014mea}. The electrons and muons which potentially originate from LLP decays are identified via a minimum requirement on the transverse distances of their tracks from the primary interaction vertex, \ie\ their respective transverse impact parameters $d_0$ (a more reliable proxy for displacement than reconstructing vertices).

The selected events are categorised into three signal regions, depending on the values of the electron and muon impact parameters. No significant excess over background was observed, so the results were interpreted as limits on the \textit{displaced supersymmetry} mo\-del~\cite{Graham:2012th}. This model contains  $R$-parity-violating stop decays ${\tilde{t}_1\to b\ell}$ with ${\ell=e,\mu,\tau}$ which happen with equal branching ratio of $\frac{1}{3}$ for all lepton flavours once lepton universality is assumed (\ie\ if the vector $\mu_L$ has three identical components). Since the stops are only produced in pairs, each signal event would have two displaced leptons. 

An implementation of this analysis \cite{Fuks:2018yku,C6B9V3_2021} is already part of the {\delphes}-based Public Analysis Database~\cite{Dumont:2014tja} of \madanalysis. It was subsequently implemented in the LLP extension of the \checkmate program~\cite{Desai:2021jsa}. It is therefore a logical choice for the first LLP analysis to implement in the \madanalysis SFS framework, which we present here, along with several improvements regarding the previous work. While the selection criteria are rather straightforward to apply, the major difficulty in this search is the lack of information about the lepton detection efficiencies for highly displaced tracks. Since the CMS Collaboration did not provide publicly available auxiliary material for this particular analysis, we have to rely on information related to the superseded analysis \cite{Khachatryan:2014mea}, available online.\footnote{See the webpage \url{https://twiki.cern.ch/twiki/bin/view/CMSPublic/DisplacedSusyParametrisationStudyForUser}.}

Recently, \cite{CMS:2016isf} was itself superseded by a full run 2 analysis \cite{CMS:2021kdm}. The updated search contains many more signal regions but no new recasting material; we attempted a naive recast but the results were not satisfactory, so we are corresponding with the convenors to improve the results and hope to return to this in future work.  

\subsection{Event selection}
The analysis targets events with exactly one electron and one muon in the final state, which are clearly identified as such and which fulfil the following preselection cuts. The imposed lower bounds on the lepton $p_T$ values are different for electrons ($\SI{42}{GeV}$) and muons ($\SI{40}{GeV}$). In both cases, the absolute value of the pseudorapidity $|\eta|$ must not exceed $2.4$. Furthermore, electrons in the overlap region of the barrel and endcap detectors ($|\eta| \in [1.44, 1.56]$) are rejected due to lower detector performance compared with other detector regions. In addition, the two leptons must satisfy simple isolation conditions. These impose an upper limit on an isolation variable $p_T^{\text{iso}}$ defined as the scalar sum of the transverse momentum of all reconstructed objects lying within a cone of a specified size $\Delta R$ and that is centred on the lepton's momentum. The limit on the isolation variable is fixed relative to the lepton momentum, and also depends on $|\eta|$ in the case of electrons, since the associated detector performance is different in the barrel and the endcaps. A summary of the criteria imposed on the electron and muon candidates considered in the event preselection, including the relevant values for the isolation cone, is given in table~\ref{CMS_event_preselection_table}.

\renewcommand{\arraystretch}{1.3}
\begin{table}\centering
  \begin{tabular}{l|c c}
    &\textbf{Electrons} & \textbf{Muons}\\\hline\hline
        \textbf{Transverse momentum}&$p_T>\SI{42}{GeV}$&$p_T>\SI{40}{GeV}$\\\hline
    &$|\eta|<1.44$ &\\
    \textbf{Pseudorapidity}&\textbf{or} &$|\eta|<2.4$\\
    &$1.56<|\eta|<2.4$&\\\hline

    \textbf{Isolation cone}&$\Delta R<0.3$&$\Delta R<0.4$\\
    \textbf{Isolation variable}&&\\
    \quad Barrel ($|\eta|<1.44$)&$p_T^{\text{iso}}<0.035\,p_T$&\multirow{2}{*}{$p_T^{\text{iso}}<0.15\,p_T$}\\
    \quad Endcaps ($|\eta|>1.56$)&$p_T^{\text{iso}}<0.065\,p_T$&\\
  \end{tabular}
  \caption{Summary of the preselection criteria imposed on the signal electrons and muons.\label{CMS_event_preselection_table}}
\end{table}
\renewcommand{\arraystretch}{1}

The analysis selection then enforces the condition that the two leptons carry an opposite electric charge and that their momenta are  separated by ${\Delta R>0.5}$. Besides these criteria, which are given explicitly in the analysis summary, we impose the condition that the position of the production vertex $\vec{v}_{\text{prod}}=(v_x,v_y,v_z)$ of each lepton satisfies
\begin{equation}
  v_0=\sqrt{v_x^2+v_y^2}<\SI{10}{cm}\,,\qquad
  v_z<\SI{30}{cm}\,.
\end{equation}
Those conditions are adopted from the superseded analysis~\cite{Khachatryan:2014mea}, for which the information was originally made available in the auxiliary material. The latter however refers to these criteria with an upper bound on $v_0$ of \SI{4}{cm}. Even though these requirements are not explicitly justified, and are not even mentioned in the publication associated with either of the analyses, it is likely that they are related to the reconstruction performance of the CMS tracker. Therefore, we decided to include them in our implementation, but with a higher upper limit on $v_0$. This is motivated by the design of the newer analysis, which is supposed to probe values of $|d_0|$ up to \SI{10}{cm}.

The analysis relies on three different signal regions (SR), which classify events according to the smallest of the two absolute values of the electron and muon impact parameters $|d_{0,e}|$  and $|d_{0,\mu}|$. The lower bound is \SI{200}{\mu m} for SR I, \SI{500}{\mu m} for SR II and \SI{1000}{\mu m} for SR III. Those regions are moreover exclusive, \ie\ there is no overlap between them. These criteria are summarised in table~\ref{cms_signal_regions_table}.

\renewcommand{\arraystretch}{1.3}
\begin{table}\centering
  \begin{tabular}{c|c}
    \textbf{Signal region}&\textbf{Defining cuts}\\\hline\hline
    \multirow{2}{*}{\textbf{SR I}}&$|d_{0,\ell}|>\SI{200}{\mu m}$ for $\ell=e,\mu$\\
                                               &$|d_{0,e}|<\SI{500}{\mu m}$\quad\textbf{\underline{or}}\quad$|d_{0,\mu}|<\SI{500}{\mu m}$\\\hline
    \multirow{2}{*}{\textbf{SR II}}&$|d_{0,\ell}|>\SI{500}{\mu m}$ for $\ell=e,\mu$\\
                                               &$|d_{0,e}|<\SI{1000}{\mu m}$\quad\textbf{\underline{or}}\quad$|d_{0,\mu}|<\SI{1000}{\mu m}$\\\hline
    \multirow{2}{*}{\textbf{SR III}}&\multirow{2}{*}{$|d_{0,\ell}|>\SI{1000}{\mu m}$ for $\ell=e,\mu$}\\
                                               &\\
  \end{tabular}
  \caption{Definition of the signal regions conforming to table~1 of ref.~\cite{CMS:2016isf}. In addition to the criteria listed explicitly, all regions respect a $|d_0|$ upper cut of \SI{10}{cm} for both leptons.\label{cms_signal_regions_table}}
\end{table}
\renewcommand{\arraystretch}{1}

\subsection{Reconstruction efficiencies\label{cms_efficiencies_sec}}
As mentioned above, no dedicated reconstruction efficiencies are provided for this analysis, but they were made available for the superseded 8 TeV analysis. We therefore use them. They are parameterised in terms of the electron and muon $|d_0|$, and separate selection efficiencies are also given in terms of their $p_T$. The recasting material for the older analysis~\cite{Khachatryan:2014mea,1317640} thus instructs us to convolve four efficiencies (two for each lepton) and to also include an additional factor of $0.95$ to account for the trigger efficiency. This is what has been done in the \checkmate implementation~\cite{Desai:2021jsa} of the CMS-EXO-16-022 search. However, there are three issues.
\begin{itemize}
\item[$\bullet$] It is easier to implement an efficiency function in \madanalysis than histogrammed data. Therefore we use a polynomial fit.
\item[$\bullet$] The efficiencies in $|d_0|$ only extend to $2\ \SI{}{cm}$, whereas the signal regions extend to $10\ \SI{}{cm}$.
\item[$\bullet$] The instructions in the auxiliary material of the analysis~\cite{Khachatryan:2014mea} direct us to impose the cuts of the (older) analysis, but to ignore lepton \emph{isolation}. In addition, we are supposed to cheat and insist that the truth-level leptons come from stop decays. Therefore, the efficiencies provided as a function of $p_T$ (which vary from about $80\%$ at the lower $p_T$ values accepted by the analysis up to about $90\%$ at higher $p_T$) convolve the effects of isolation and some of the cuts, and are in addition model-specific.
\end{itemize}

\begin{figure}\centering
  \includegraphics[width=\columnwidth]{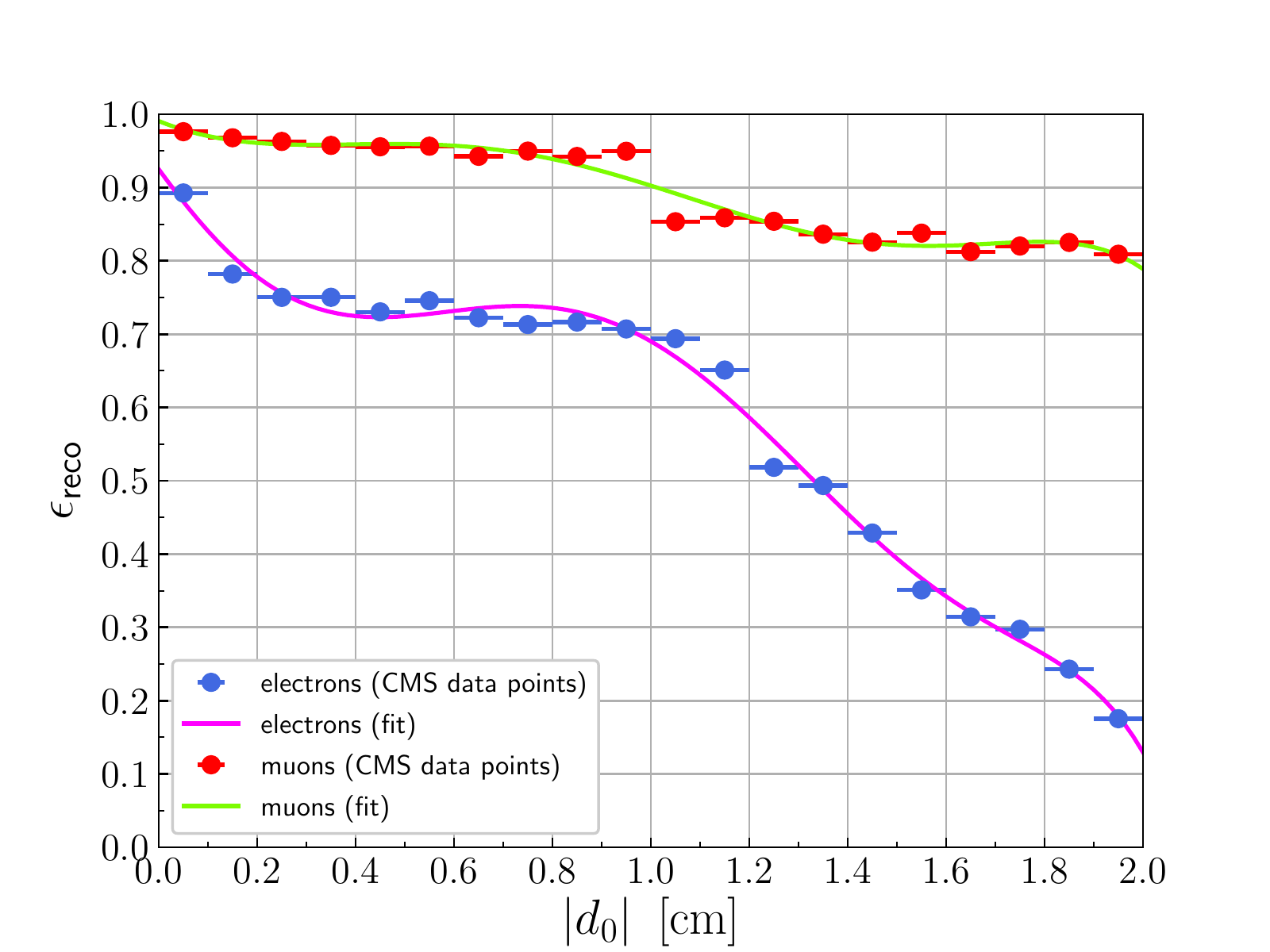}
  \caption{Electron and muon reconstruction efficiencies as extracted from the superseded CMS analysis~\cite{Khachatryan:2014mea}. The plot was generated with data files from the corresponding {\sc HEPData} entry~\cite{1317640} and the polynomial fits of eq.~\eqref{eq_d0_efficiency_fit} generated by the author of \cite{C6B9V3_2021}.\label{d0_efficiencies_figure}
  }
\end{figure}

The implementation in \cite{Desai:2021jsa} took the approach of using \delphes, ignoring lepton isolation and relying on the efficiencies provided by the CMS Collaboration. We tried this but without insisting on selecting only leptons issued from top squark decays. We found poor agreement with the CMS analysis, notably for small LLP lifetimes. On the other hand, once we imposed lepton isolation directly in our implementation then the combined effect was too aggressive. Therefore, we chose to omit the $p_T$-dependent efficiencies (recall that there are four functions, two each depending on $|d_0|$ for the electron and muon, and two depending on $p_T$ of the electron and muon, respectively) and to keep lepton isolation at the level of the analysis, rendering our code (hopefully) more model-independent: we believe that the information in the $p_T$ functions must essentially be the cuts relating to isolation requirements. The polynomial fits of the efficiencies for $|d_0|<\SI{2}{cm}$ were realised by the author of \cite{C6B9V3_2021}, which we also employ. This gives
\be\bsp
  \epsilon_{\text{reco},e}&(x) = 0.924921 - 0.917957 x + 0.522007 x^2 \\
     & + 2.87189 x^3 - 4.9321 x^4 + 2.72756 x^5\\
     & - 0.506107 x^6,\\
   \epsilon_{\text{reco},\mu}&(x)=
    0.99067 - 0.271852 x + 0.743217 x^2\\
   & - 0.611108 x^3  - 0.260292 x^4 + 0.423266 x^5 \\
   &- 0.111279 x^6,
\esp\label{eq_d0_efficiency_fit}\ee
for electrons and muons, respectively, with $x \equiv |d_0|/\SI{}{cm}$. The tabulated efficiencies and the resulting fits are shown in figure~\ref{d0_efficiencies_figure}.

Above transverse impact parameters of \SI{2}{cm} we must make a choice due to the absence of experimental data. That made in the \checkmate implementation of ref.~\cite{Desai:2021jsa} is to provide one bin and make a $\chi^2$ fit to minimise the differences at the level of the cutflows. This leads to an electron efficiency of $0.06$ and a muon efficiency of $0.01$. Especially for muons this is effectively consistent with zero, and does not seem reasonable given that at \SI{2}{cm} the efficiencies are much higher. We can only speculate as to how/why that choice and also some other choices were made in ref.~\cite{Desai:2021jsa}. In fact, in the SFS analysis we typically found that we were cutting too many events compared with the experimental exclusion. So instead we investigated having either a constant efficiency equal to the value at \SI{2}{cm}, or a linear extrapolation. It is the latter that produced a better agreement, and so the one we adopt. We thus assume that the lepton efficiencies decrease linearly to zero at $|d_0|=\SI{10}{cm}$, which is the maximum value at which they can be selected in the analysis. The efficiencies for $\SI{2}{cm}<|d_0|\leq\SI{10}{cm}$ read
\be\bsp
  \epsilon_{\text{reco},e}(x)  & =0.15 - 0.01875 \times \left(x - 2\right), \\
  \epsilon_{\text{reco},\mu}(x)& =0.8 - 0.1      \times \left(x - 2\right),
\esp\ee
still with $x \equiv |d_0|/\SI{}{cm}$, and they vanish for any $|d_0|$ value larger than \SI{10}{cm}.

\subsection{Validation of our implementation\label{cms_validation_sec}}
The signal process considered in the CMS-EXO-16-022 analysis consists of the production of a pair of long-lived top squarks that then decay into an electron-muon pair,
\be pp\to \tilde{t}\tilde t^* \to \to b \ell^+\ \bar b \ell^{\prime-},\label{eq:procvalcms}\ee
with $\ell, \ell' =e,\mu$. The only validation material available for this process is related to four scenarios in which the top squark mass is fixed to $m_{\tilde{t}}= \SI{700}{GeV}$, and which differ by the stop decay length that ranges from 0.1 to \SI{100}{cm}. In each case, the CMS Collaboration provides the number of signal events populating the three signal regions of the analysis (see table 4 of ref.~\cite{CMS:2016isf}).

To reproduce the CMS results, we generate four samples of 400,000 events with \pythia~\cite{Sjostrand:2014zea}. In our simulation chain, we focus on the process~\eqref{eq:procvalcms} and enforce the decays $\tilde{t}\to b \ell$ to have equal branching ratios of 1/3 for each lepton flavour. It is important to stress that we do not impose the condition that the stops decay only into electrons and muons, but we also consider stop decays into taus. This last subprocess contributes differently to the signal. The resulting electrons and muons indeed exhibit different properties (like their $p_T$) that cannot be modelled through a simple multiplicative factor applied to signal yields as obtained after restricting the stops to decay solely into electrons and muons. For a fair comparison with the CMS experimental results and simulations, we next rescale the obtained number of events to match a stop pair-production cross section evaluated at the next-to-leading-order and next-to-leading-logarithmic (NLO+NLL) accuracy in quantum chromodynamics (QCD)~\cite{Beenakker:2010nq}, \ie\ a cross section of ${\sigma\approx\SI{67.05}{fb}}$.

\renewcommand{\arraystretch}{1.3}
\begin{table}\centering
  \begin{tabular}{c|c|c|c}
    {\boldmath $c\tau$}&\textbf{SR I}&\textbf{SR II}&\textbf{SR III}\\
    {\boldmath $\mathrm{[cm]}$}&{\color{blue}CMS} / MA5&{\color{blue}CMS} / MA5&{\color{blue}CMS} / MA5\\\hline\hline
                       &{\color{blue}$3.8\pm0.2$}&{\color{blue}$0.94\pm0.06$}&{\color{blue}$0.16\pm0.02$}\\
    \textbf{0.1}&$2.9$&$0.74$&$0.15$\\
                       &{\color{red}$\delta_\epsilon=\SI{23.0}{\percent}$}&{\color{red}$\delta_\epsilon=\SI{21.6}{\percent}$}&${\color{red}\delta_\epsilon=\SI{6.6}{\percent}}$\\\hline
                       &{\color{blue}$5.2\pm0.4$}&{\color{blue}$4.1\pm0.3$}&{\color{blue}$7.0\pm0.3$}\\
    \textbf{1}&$3.8$&$3.4$&$5.7$\\
                    &{\color{red}$\delta_\epsilon=\SI{26.1}{\percent}$}&{\color{red}$\delta_\epsilon=\SI{17.7}{\percent}$}&{\color{red}$\delta_\epsilon=\SI{18.4}{\percent}$}\\\hline
                       &{\color{blue}$0.8\pm0.1$}&{\color{blue}$1.0\pm0.1$}&{\color{blue}$5.8\pm0.2$}\\
    \textbf{10}&$0.65$&$0.82$&$4.6$\\
                       &{\color{red}$\delta_\epsilon=\SI{18.6}{\percent}$}&{\color{red}$\delta_\epsilon=\SI{17.8}{\percent}$}&{\color{red}$\delta_\epsilon=\SI{20.5}{\percent}$}\\\hline
                       &{\color{blue}$0.009\pm0.005$}&{\color{blue}$0.03\pm0.01$}&{\color{blue}$0.27\pm0.03$}\\
    \textbf{100}&$0.023$&$0.037$&$0.22$\\
                       &{\color{red}$\delta_\epsilon=\SI{153.0}{\percent}$}&{\color{red}$\delta_\epsilon=\SI{24.2}{\percent}$}&{\color{red}$\delta_\epsilon=\SI{18.6}{\percent}$}\\
  \end{tabular}
  \caption{Expected number of events passing the selection criteria defining each of the CMS-EXO-16-022 signal regions, as obtained with our SFS implementation (black). We compare our predictions with the official CMS results (blue), and show the relative deviation $\delta_\epsilon$ in the selection efficiency (red). \label{tab:results_cms_sfs}
  }
\end{table}
\renewcommand{\arraystretch}{1}

Our results are shown in table~\ref{tab:results_cms_sfs} and demonstrate excellent agreement for all considered lifetimes and signal regions, with as a possible exception the SR I region for $c\tau = \SI{100}{cm}$. Here, a deviation reaching $150\%$ is found. This is however irrelevant, as the corresponding efficiency is tiny and the uncertainty on the CMS result is larger than $50\%$ (which is due in particular to the associated low statistics). Moreover, we have no information on the large numerical errors potentially plaguing the CMS predictions.

To complete the validation of our implementation, we derive an exclusion contour at 95\% confidence level in the $(m_{\tilde t}, c\tau)$ plane, and compare our findings with the exclusion plot included in the CMS analysis publication. Such a comparison is complicated by two factors.
\begin{itemize}
\item[$\bullet$] The analysis provides data-driven background estimates for each signal region \emph{without uncertainties}, giving instead just a \emph{maximum} value for the background yields.
\item[$\bullet$] The exclusion associated with scenarios featuring an intermediate stop lifetime depends significantly on a combination of the results emerging from the different signal regions, yet no correlation data are given.
\end{itemize}

The older \madanalysis analysis~\cite{C6B9V3_2021} took the approach of choosing the median value of the background with an uncertainty of $100\%$. In the \checkmate implementation of ref.~\cite{Desai:2021jsa}, the expected backgrounds are instead taken to be equal to the \emph{maximum} value possible for the background with an (arbitrary) uncertainty of $10\%$.

\begin{figure}\centering
  \includegraphics[width=0.49\textwidth]{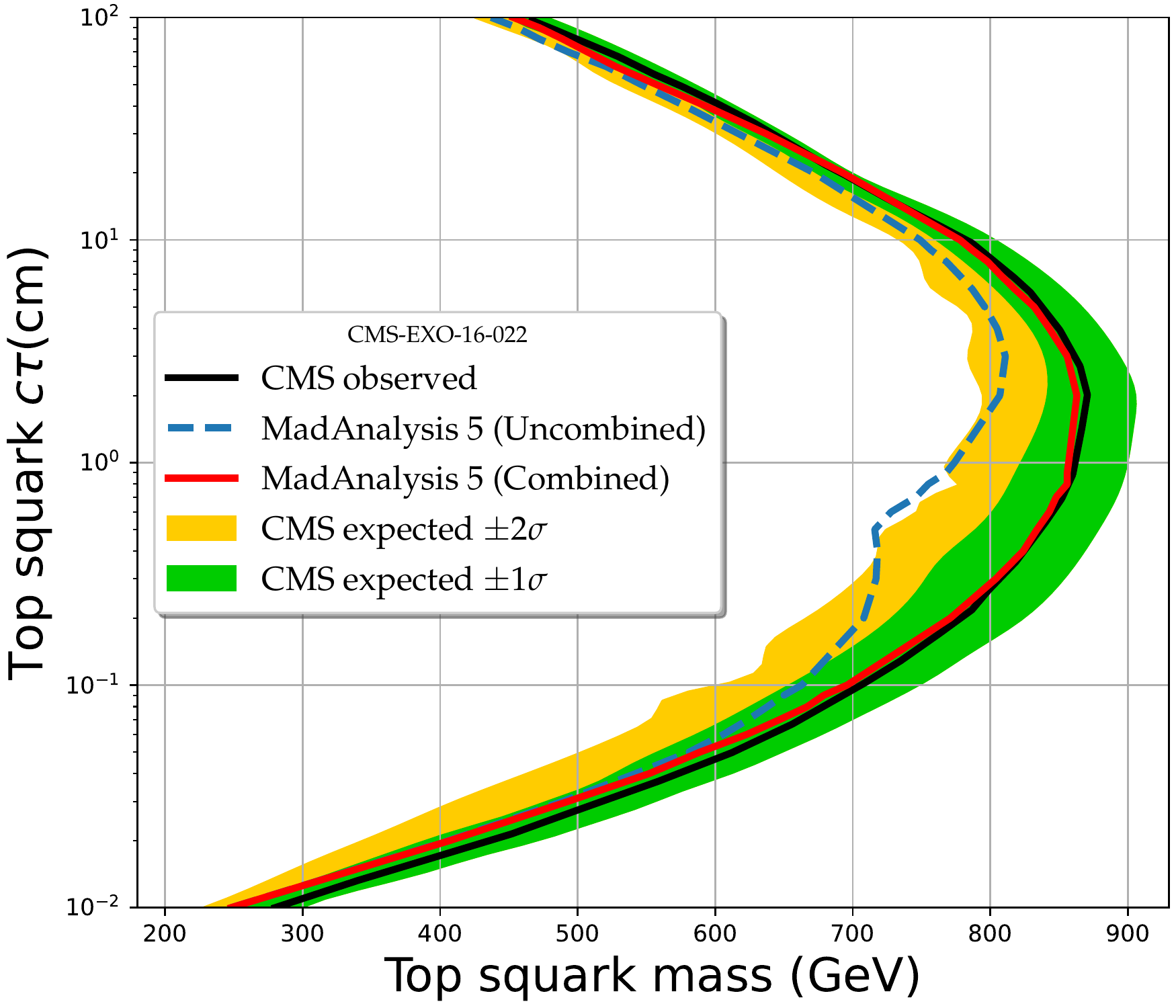}
\caption{Exclusion contours in the $(m_{\tilde t}, c\tau)$ as obtained using our {\madanalysis} (SFS) implementation (red), and the CMS official exclusion curve (black, with the $1\sigma$ and $2\sigma$ bands given in green and yellow).\label{CMS_excl_plot_repro}}
\end{figure}

In the public version of the analysis~\cite{SVILPJ_2021}, we provide the same values for the background as in the old implementation; \madanalysis will thus compute the exclusion based on choosing the ``best'' signal region according to the standard procedure, and this is the most appropriate conservative choice that can be made. However, for the purposes of reproducing the CMS exclusion curve, we also privately modified the statistics module of \madanalysis to draw the expected background counts from a flat distribution extending up to the maximum value, and combined the predictions for the different signal regions by assuming no correlations. We show both results in figure~\ref{CMS_excl_plot_repro}, with the default statistics choice labelled as ``Uncombined,'' and our modified one as ``Combined.'' The striking agreement for the combined curve over the entire considered range with the observed CMS exclusions (which we scraped digitally, since the values are not provided publicly) is a further validation of our analysis, but also a reminder of the importance for the experimental collaborations to provide correlation data. Since correlation matrices (or information on the validity of combining signal regions in an uncorrelated way) have not been provided by the CMS Collaboration, the ``Combined'' approach remains private, and only the ``Uncombined'' version is publicly available.\footnote{On the other hand, it is very straightforward for a knowledgeable user to switch between the two by modifying the associated {\tt \.info} file.}

\section{Displaced vertices with oppositely charged leptons (ATLAS-SUSY-2017-04)\label{sec_ATLAS_LLP_analysis}}
\setcounter{equation}{0}
\subsection{Generalities}
The ATLAS search \cite{ATLAS:2019fwx} targets events featuring displaced vertices (DVs) associated with a pair of leptons ($ee$, $e\mu$ or $\mu\mu$) with an invariant mass greater than \SI{12}{GeV}. It uses run 2 data samples from 2016, corresponding to an integrated luminosity of \SI{32.8}{fb^{-1}} at $\sqrt{s}=\SI{13}{TeV}$. No events with a dileptonic displaced vertex compatible with all selection criteria were observed. This analysis mainly considered $R$-parity-violating supersymmetric scenarios in which colourful scalars (squarks $\tilde{q}$) are produced and decay promptly to a long-lived heavy fermion (neutralino, $\tilde{\chi}^0_1$),
\begin{align}
p p \to \tilde q \tilde q^* \to \tilde{\chi}^0_1 q\ \tilde{\chi}^0_1 \bar q.
\label{eq:hard1}\end{align}
The neutralino next decays into a pair of opposite-sign leptons and a neutrino,
\begin{align}
\tilde{\chi}^0_1 \to \ell_k^\pm \ell_{i/j}^\mp \nu_{j/i},
\end{align}
where $i,j,k$ denote fermion generation. The branching ratios into different $i,j,k$ flavours are determined by a coupling $\lambda_{ijk},$ and the analysis focuses on two different scenarios, namely models featuring either a single $\lambda_{121}$ coupling or a single $\lambda_{122}$ coupling. The ``$\lambda_{121}$ scenario'' leads to $\mathrm{BR}(\tilde{\chi}^0_1 \rightarrow ee\nu) = \mathrm{BR}(\tilde{\chi}^0_1 \rightarrow e\mu\nu) =0.5$, while the ``$\lambda_{122}$ scenario'' leads to $\mathrm{BR}(\tilde{\chi}^0_1 \rightarrow \mu\mu\nu) = \mathrm{BR}(\tilde{\chi}^0_1 \rightarrow e\mu\nu) =0.5$. The results are also interpreted in the framework of a heavy sequential $Z'$ boson~\cite{Altarelli:1989ff}, which has thus the same couplings as the SM $Z$-boson, but with a lifetime that is artificially modified to make it long-lived. Such a long-lived $Z'$ would be excluded by searches for displaced hadronic jets, but is an excellent and simple prototype for a two-body LLP decay. It could hence be used to obtain signal selection efficiencies which can be applied to a large set of comparable models.

As opposed to the CMS analysis presented in the previous section, this analysis requires the pair of oppositely charged leptons to originate from the \emph{same} DV. It also allows for several DVs per event. Moreover, an event is not rejected automatically when a DV does not fulfil all of the requirements imposed, as long as there is at least one other DV surviving the cuts. 

An extensive amount of auxiliary material for the reinterpretation of the results of this analysis is provided on \hepdata~\cite{1745920} and on the analysis twiki page.\footnote{See the webpage \url{https://atlas.web.cern.ch/Atlas/GROUPS/PHYSICS/PAPERS/SUSY-2017-04/}.} Most importantly, the auxiliary material contains information about individual efficiencies for the vetoes that are applied on DVs in different detector regions, and the overall event selection efficiency after all cuts. Whereas these have a significant impact on the final result, they are rather model- and parameter point-dependent, so that the user must use caution in applying them.

\subsection{Selection criteria}

The event selection is composed of two types of cuts. First, a number of cuts keep or reject events as a whole. Next, the remaining cuts perform a selection of displaced vertex candidates, but do not reject events unless the number of vertex candidates is reduced to zero.

\subsubsection{Event-level requirements\label{subsec_ATLAS_event_level_requirements}}

Here we list the event requirements as we apply them in our implementation.\vspace{.15cm}

\noindent \textbf{Triggers:} We require that at least one out of three triggers, targeting muons and electrons, has fired. A \textit{muon trigger} requires a muon with ${p_T>\SI{60}{GeV}}$ and $|\eta|<1.05$. On the other hand, since electrons are hard to reconstruct, the analysis also triggers on \emph{photons} from their electromagnetic showers.  Since we do not simulate the effect of electrons passing through the calorimeter, we are more generous and apply the same triggers to generator-level electrons. Hence a \textit{single photon trigger} requires an electron or a photon with $p_T>\SI{140}{GeV}$, and a \textit{diphoton trigger} requires two photons or electrons with $p_T>\SI{50}{GeV}$.\vspace{.15cm}

\renewcommand{\arraystretch}{1.3}
\setlength\dashlinedash{2pt}
\addtolength{\tabcolsep}{-4.2pt}
\begin{table}\scriptsize
  \begin{tabular}{c|lccc|lccc}
    \textbf{Trigger} & \multicolumn{4}{c|}{\textbf{Candidate 1}} & \multicolumn{4}{c}{\textbf{Candidate 2}}\\
                     & &{\boldmath $p_T$}&{\boldmath $|\eta|$}&{\boldmath $|d_0|$}&&{\boldmath $p_T$}&{\boldmath $|\eta|$} & {\boldmath $|d_0|$}\\
                         & &{$\mathrm{[GeV]}$}&&{$[\mathrm{mm}]$}&&{$[\mathrm{GeV}]$}& & {$[\mathrm{mm}]$}\\\hline\hline
    {\boldmath $\mu$}&$\mu$&$>62$&$<1.07$&$>1.5$ &\multicolumn{4}{c}{no candidate 2 required}\\\hline
    \multirow{4}{*}{{\boldmath $\gamma$}}&&&&&$\gamma$&$>10$&$<2.5$&------------\\
                     &$\gamma$&$>150$&$<2.5$&------------&$e$&$>10$&$<2.5$&$>2.0$\\
                     &&&&&$\mu$&$>10$&$<2.5$&$>1.5$ \\\cdashline{2-9}
                     &$e$&$>150$&$<2.5$&$>2.0$&\multicolumn{4}{c}{no candidate 2 required}\\\hline
    \multirow{4}{*}{{\boldmath $\gamma\gamma$}}&$\gamma$&$>55$&$<2.5$&------------&$\gamma$&$>55$&$<2.5$&------------\\\cdashline{2-9}
                     &\multirow{2}{*}{$e$}&\multirow{2}{*}{$>55$}&\multirow{2}{*}{$<2.5$}&\multirow{2}{*}{$>2.0$}&$e$&$>55$&$<2.5$&$>2.0$\\
                     &&&&&$\gamma$&$>55$&$<2.5$&------------\\\cdashline{2-9}
                     &$e$ \textbf{\boldmath($*$)}&$>55$&$<2.5$&------------&$\gamma$&$>55$&$<2.5$&------------\\
  \end{tabular}
  \caption{Preselection criteria based on table 1 of ref. \cite{ATLAS:2019fwx}. According to the triggers which have been
    fired beforehand, the presence of one or two candidate objects with the specified properties is required
    for at least one of the triggers. As explained in the text, electron candidates tagged with a star (*) must be \textit{loosely isolated}.
    \label{table_ATLAS_preselection}}
\end{table}
\renewcommand{\arraystretch}{1}

\noindent \textbf{Preselection:} Preselection requirements are then applied on the one or two candidate particles that emerge from the triggers (and that could be electrons, muons or photons). The constraints applied on their $p_T$, $\eta$, $d_0$ and isolation are listed in  table~\ref{table_ATLAS_preselection}. Electrons must have an impact parameter $|d_0|>\SI{2}{mm}$ unless they satisfy the \textit{loose electron} isolation criteria, whereas photons are not associated with tracks so that they cannot be subject to any $d_0$ requirement. They are instead \emph{all} required to be loosely isolated. Two different isolation criteria must be fulfilled by loose electrons. Calorimeter isolation requires that the sum $E_T^{\text{cone20}}$ of the energy deposits inside a cone of fixed size $\Delta R<0.2$ around each electron is as small as
\begin{equation}
    \frac{E_T^{\text{cone20}}}{p_T}<0.2\,.
\end{equation}
Track isolation restricts the scalar sum $p_T^{\text{varcone}}$ of the $p_T$ of all other tracks in a cone with variable size $\Delta R$ around the electron track to fulfil
\be\bsp
 \frac{p_T^{\text{varcone}}}{p_T}<0.15\
      \text{with}\ \Delta R=\min\left(\frac{10~{\rm GeV}}{p_T}, 0.2\right)\,.
\esp\ee
For photons, we enforce instead that
\begin{equation}
     \frac{E_T^{\text{cone20}}}{p_T}<0.065, \qquad \frac{p_T^{\text{cone20}}}{p_T}< 0.05,
\end{equation}
with a constant cone size of $\Delta R = 0.2$. For muons, we require a displacement of $|d_0|>\SI{1.5}{mm}$. Whereas in the ATLAS analysis this requirement is skipped if the muon track is poorly reconstructed, we simply apply it to all muons in our \madanalysis implementation.\vspace{.15cm}

\noindent \textbf{Cosmic ray veto:} We reject events compatible with the presence of cosmic ray muons, \ie\ events in which \emph{any} of the possible lepton pairs (in case electrons are reconstructed as muons) with pseudorapidities $\eta_{1,2}$ and polar angles $\phi_{1,2}$ satisfy
\begin{equation}
    \Delta R_{\text{cos}}\equiv\sqrt{(|\phi_1 - \phi_2| - \pi)^2 + (\eta_1 + \eta_2)^2}<0.01\,. 
\end{equation}

\noindent \textbf{Primary vertex:} After the trigger selections, the analysis requires the presence of a \emph{primary vertex}. This is clearly redundant for an SFS-based analysis of only signal events. We therefore do not implement this cut.\vspace{.15cm}

\noindent \textbf{Displaced vertex:} The set of event-level requirements ends by enforcing events to contain tracks which form at least one displaced vertex. To allow for highly displaced vertices, the ATLAS \textit{standard tracking} algorithm is supplemented with a \textit{large radius tracking} algorithm \cite{ATLAS:2017zsd}, which significantly relaxes the requirements imposed on the tracks. For example, it raises the $|d_0|$  upper limit from 10 to \SI{300}{mm} and the $|d_z|$ upper limit from 250 to \SI{1500}{mm}. The positions of the displaced vertices are obtained from the reconstructed tracks with a \textit{vertexing} algorithm~\cite{Aaboud:2017iio}, via the successive combination of intersecting tracks (taking into account the uncertainties associated with their reconstruction) to vertices and the merging of vertices when their distance is small enough. However, our simulated events already contain information about the vertex positions (which can be slightly modified if the magnetic field is taken into account by using the SFS particle propagator module introduced in section~\ref{sec:ma5}). Therefore it would be overkill to attempt to implement the ATLAS algorithm, and fruitless since we have insufficient information regarding the tracker efficiency. Instead, we simply generate an object representing a displaced vertex for each external final-state particle compatible with the track requirements for the DV reconstruction, \ie\ tracks satisfying
\be\bsp
    p_T>\SI{1}{GeV}\,,\ \\ \SI{2}{mm}<|d_0|<\SI{300}{mm}\,,\ \quad |d_z|<\SI{1500}{mm}\,. 
\esp\ee
Then a very simple merging is performed. It consists of replacing two DV objects separated by a distance smaller than \SI{1}{mm} by a single DV object and assigning all particles associated with the initial vertices to the new DV object. The position assigned to the new DV object is arbitrarily set to the position of one of the two
vertices. This should reproduce the merging procedure well enough, assuming that the LLP decay products are either stable or promptly decaying, or sufficiently short-lived particles. In this way, there is no risk to end up with two displaced vertices when there should be only one.

\subsubsection{Vertex-level requirements\label{subsec_atlas_vertex_level_requirements}}

In a second step, the analysis strategy targets the ensemble of displaced vertices in a given event. Individual (displaced) vertices are rejected if they fail one of the following requirements.\vspace{.15cm}

\noindent \textbf{Vertex fit:} The ATLAS analysis requires a good fit for the vertex; similarly to the primary vertex cut, we cannot reproduce the algorithm employed in the analysis. However, the \hepdata entry provides per-decay acceptances, intended to be applied after all previous cuts. In our \madanalysis SFS implementation we use them as vertex reconstruction efficiencies, as discussed further in section~\ref{subsec_atlas_efficiencies}.\vspace{.15cm}

\noindent \textbf{Transverse displacement:} Prompt decays and decays after a small displacement are avoided by considering only DVs at positions ${\bf x}_{\text{DV}}=(x_{\text{DV}},y_{\text{DV}},z_{\text{DV}})$ for which the transverse displacement from the proton collision axis is larger than \SI{2}{mm}. As in any given simulated Monte Carlo event protons are assumed to collide at the origin, this gives
\begin{equation}
    r_{xy}\equiv\sqrt{x_{\text{DV}}^2+y_{\text{DV}}^2}>\SI{2}{mm}\,.
\end{equation}

\noindent\textbf{Fiducial volume:} Only vertices in a restricted detector volume, where track and vertex reconstruction are expected to be reliable, are taken into account. This \textit{fiducial volume} is a cylinder around the beam axis of radius of \SI{300}{mm} and length of \SI{600}{mm}. We thus consider only DVs with a position satisfying
\begin{equation}
  r_{xy}<\SI{300}{mm}\,,\qquad |z_{\text{DV}}|<\SI{300}{mm}\,.
\end{equation}

\noindent\textbf{Material veto:} The ATLAS detector itself contributes to the presence of displaced vertices from interactions of primary particles with detector material~\cite{Aaboud:2017iio,Aaboud:2016poq,Aaboud:2017pjd}. The \emph{material veto} removes electrons originating from within the tracking layers or support structures, which consists of about $42\%$ of the fiducial volume. This would be an exceedingly difficult requirement to simulate in the SFS framework. Fortunately, the ATLAS Collaboration has provided an efficiency map splitting the fiducial volume into a grid in $r_{xy}$ and $z_{\text{DV}}$, averaged over the polar angle $\phi$. This map gives the \emph{fraction} of each grid element eliminated by the veto. Our ATLAS-SUSY-2017-04 implementation in {\madanalysis} hard-codes this map into a C++ function (instead of a fitting function given via the analysis card) and randomly discards vertices with the probability given by the result.\vspace{.15cm}

\noindent\textbf{Disabled pixel modules veto:} Similar to the material veto, DVs localised in front of disabled pixel modules are vetoed. This corresponds to $2.3\%$ of the fiducial volume. Efficiencies averaged over the polar angle are again provided in terms of $r_{xy}$ and $z_{\text{DV}}$ by the ATLAS Collaboration. We implement them through a hard-coded function.\vspace{.15cm}

\noindent\textbf{Two leptonic tracks:} At least two leptonic tracks must be associated with each reconstructed vertex. In the provided cutflows (and hence in our implementation) these are implemented as successive cuts on the number of leptons associated with each vertex, $N(\ell) \ge 1$ and $N(\ell) \ge 2.$.\vspace{.15cm}

\noindent\textbf{Invariant mass:} The invariant mass of the sum of the momenta of the tracks associated with the displaced vertices give a lower bound on the mass of the decaying long-lived particle. It is required to exceed \SI{12}{GeV}.\vspace{.15cm}

\noindent\textbf{Trigger and preselection matching:} The trigger and preselection criteria are also required to hold for the subset of particles associated with each of the displaced vertices. This ensures that the energetic leptons and photons considered originate from the displaced vertices themselves. Naively for typical signal events targeted by the analysis, this should be of negligible impact (and indeed this is generally what we observe). In contrast, for the decay of a relatively light LLP we shall see from the cutflows that this could be very significant, presumably because of triggered photons (and possibly leptons) stemming from initial-state or final-state radiation.\vspace{.15cm}

\noindent\textbf{Oppositely charged lepton pair:} Each displaced vertex is enforced to involve at least one pair of leptons ($ee$, $\mu\mu$ or $e\mu$) with opposite electric charge.

\subsection{Efficiencies\label{subsec_atlas_efficiencies}}
As described above, three types of efficiencies are provided in the \hepdata repository for this analysis. In the auxiliary file describing the recasting material, the user is moreover instructed to apply essentially the cuts described above and then compute a weight for each vertex based on the material veto and disabled pixel veto efficiencies. This yields a per-decay \emph{acceptance}, and next the per-decay \emph{efficiency} can be computed by using the provided maps.

The \emph{per-event} efficiency can be computed by a simple formula. As already mentioned, in our \madanalysis implementation we do not apply a reweighting but instead randomly eliminate vertices with probability given by the veto weight. Furthermore, we apply the per-decay efficiency as a vertex reconstruction efficiency. This is somewhat simpler to implement (at the expense of requiring more statistics) and obviates the need for a formula to combine multiple vertices per event, but should otherwise be equivalent.

While the material and disabled pixel veto weights are expected to be largely model-independent (except for perhaps models having a large polar angle dependence with multiple DVs per event), the efficiency information convolves trigger/preselection and vertex reconstruction efficiencies, and so is highly model-de\-pen\-dent. The \hepdata material contains sets of bin\-ned efficiencies in two or three variables, for two models and several masses.

The first model is a toy sequential $Z'$ model in which the extra gauge boson has the same couplings as the SM $Z$ boson, but a different mass. Such a model would be excluded by searches for dijet resonances, but the results/efficiencies can be applied to any model having events with a single LLP decaying to a lepton pair. Efficiencies are given for fixed masses $m(Z')$ of 100, 250, 500, 750 and \SI{1000}{GeV}, with a binning in the transverse distance $r_{xy}$ and the lepton pair transverse momentum $p_T(\ell\ell')$.

The second model is an $R$-parity-violating supersymmetric (RPV SUSY) model where a heavy fermion (neutralino) decays to two leptons and a neutrino.  Obviously, this has implications for the kinematics of the pair of visible leptons, as they do not have to conserve the four-momentum of the neutralino. For this reason, the mass of the neutralino cannot be determined directly from the momenta of the visible lepton pair. Unlike the $Z'$-efficiencies, which depend on the mass of the long-lived $Z'$ but are only given for fixed $m(Z')$ values, the RPV SUSY efficiencies depend on the invariant mass of the lepton pair $m(\ell\ell')$, without any assumption on the neutralino mass. They are therefore binned in the three variables $r_{xy}$, $p_T(\ell\ell')$ and $m(\ell\ell')$.

In our implementation we have implemented all of the available efficiencies \emph{in two versions}, one for two-body LLP decays and one for three-body decays. The user must therefore choose the version appropriate for their model; we tested the effect on the $Z'$ model using the RPV efficiencies and found deviations of about \SI{15}{\percent}. However, they should still exercise some caution. Firstly, there can be models with the same signature but different kinematics, for example when the invisible particle escaping a vertex is massive, instead of a massless neutrino; in such cases there are no available efficiencies. Secondly, we also do not know how much of the information corresponds to trigger/preselection efficiencies as compared with vertex reconstruction. Consequently, if the production mechanism changes, there would also be an unknown (but most likely of the order of tens of percent) induced error.

\subsection{Validation\label{subsec_ATLAS_validation}}

A significant amount of material is provided both in the analysis note and as auxiliary online information, and can be used for the validation of the implementation.  This includes cutflows, plots (with tabulated data on \hepdata) of per-decay and per-event signal efficiencies and excluded cross sections. It should therefore be expected that the final efficiencies should be almost exactly reproducible for the example benchmarks at least.

For the $Z'$ toy model, there are cutflows provided for six \emph{different event samples}, in which the $Z'$ boson has a mass of either \SI{100}{GeV} or \SI{1000}{GeV}, and decays exclusively into one of the three dilepton final states $ee$, $e\mu$ or $\mu\mu$.

For the RPV SUSY model, the cutflows are provided for event samples combining the $\lambda_{121}$ and $\lambda_{122}$ scenarios, where $\mathrm{BR}(\tilde{\chi}_1^0 \rightarrow ee\nu) = \mathrm{BR}(\tilde{\chi}_1^0 \rightarrow e\mu\nu) = \mathrm{BR}(\tilde{\chi}_1^0 \rightarrow \mu\mu\nu) = \nicefrac{1}{3}$. In analogy to the $Z'$ cutflows, the material contains separate cutflows for each of the neutralino decay modes (in a $ee$, $e\mu$ and $\mu\mu$ final state), and an event is counted in the cutflow if it contains at least one vertex associated with the corresponding lepton pair. Here, two different neutralino lifetimes are considered for a single configuration of the squark and neutralino masses, which leads to six different cutflows as well. In addition to the cutflows, plots of selection efficiencies varying the LLP lifetime were provided in the analysis paper for several squark and neutralino masses. To validate our analysis, we performed a scan for the RPV SUSY model for two configurations of the quark and neutralino masses over a wide spectrum of neutralino lifetimes, considering the $\lambda_{121}$ and $\lambda_{122}$ couplings separately. The parton-showered and hadronised samples were then passed to our {\madanalysis} implementation of the analysis to determine the selection efficiencies, including the effects of the detector through the SFS framework. The code is available publicly~\cite{31JVGJ_2021} and has been used with the goal of reproducing some of the ATLAS results presented in the figures 3-5 of ref.~\cite{ATLAS:2019fwx}.

\subsubsection{Event generation}

In the light of the available cutflows, we made use of \pythia (v8.244)~\cite{Sjostrand:2014zea} to generate six separate samples of 20,000 events describing the production of a sequential $Z'$ boson and the appropriate $ee$, $e\mu$ or $\mu\mu$ decays for \emph{both} $m(Z')=$\SI{100}{GeV} and \SI{1000}{GeV}. In all cases the proper decay length was fixed at $c\tau = \SI{250}{mm}$. These samples could be directly passed to \madanalysis.

Various samples of 20,000 events associated with long-lived neutralino production in the RPV SUSY mo\-del were generated with the {\madgraph}~\cite{Alwall:2014hca} package (v2.8.3.2), which we used in conjunction with \pythia (v8.244) for parton showering and hadronisation. The hard-scattering process~\eqref{eq:hard1} is simulated by merging\footnote{The number of events in the sample is reduced to some extent due to the application of the MLM matching scheme, leading to a varying number of events for the different benchmark points. All event numbers given refer to the {\it initially} generated number of events with {\madgraph} before matching. The same is true in the following for any samples generated with {\madgraph} and {\pythia}, and in particular for the vector-like lepton model studied in section~\ref{sec_VLL_LLP}.} matrix elements containing up to two additional partons via the multi-leg merging (MLM) scheme~\cite{Mangano:2001xp,Mangano:2006rw,Alwall:2008qv} as modified by the internal \madgraph interface to \pythia. Those matrix elements rely on the Minimal Supersymmetric Standard Model (MSSM) UFO model shipped with \madgraph~\cite{Duhr:2011se,Degrande:2011ua}, are convoluted with the LO set of NNPDF~2.3~\cite{Ball:2012cx} parton distribution functions, and include matching cuts derived from a matching scale set to of one fourth of the squark mass. The parameters defining the benchmark scenarios considered (including decay tables so that \pythia could handle squark and neutralino decays) are taken from the parameter card provided on \hepdata, so that comparisons of the analysis cutflows could be made on a cut-by-cut basis. Our validation employs two samples which combine all neutralino decay modes with equal branching ratios $\rm{Br}(\tilde{\chi}_1^0\to ee\nu)=\rm{Br}(\tilde{\chi}_1^0\to e\mu\nu) =\rm{Br}(\tilde{\chi}_1^0\to \mu\mu\nu)=\nicefrac{1}{3}$. A classification of the events is then done on the level of the analysis code, depending on the types of leptons originating from the displaced vertices present.

For the validation of the efficiencies, the couplings $\lambda_{121}$ and $\lambda_{122}$ are considered separately, and we generated samples of 50,000 events for two configurations of the masses ($m(\tilde{q})=\SI{700}{GeV}$, $m(\tilde{\chi}_1^0)=\SI{50}{GeV}$ and $m(\tilde{q})=\SI{1600}{GeV}$, $m(\tilde{\chi}_1^0)=\SI{1300}{GeV}$) and 21 different neutralino lifetimes ranging from $c\tau=\SI{1}{mm}$ to $c\tau=\SI{10000}{mm}$, for both of the $R$-parity-violating couplings (\ie\ in total 84 samples).

\subsubsection{Cutflows}
We used the cutflow tables provided for both models as guidelines during the implementation, with the intention to reproduce the efficiencies of the individual intermediate cuts of the analysis, and not just the final efficiencies. However, the per-decay efficiencies provided are not just simple reconstruction efficiencies, but fold in the preselection. Moreover, pile-up effects can induce displaced vertices along the beamline relative to the hard-scattering event, and could thus yield important effects. It was therefore quickly apparent that reproducing the cutflows in detail would not be possible.

To test the impact of the pile-up (for which there is not yet support in the SFS framework), we implemented the analysis in {\sc HackAnalysis}~\cite{Goodsell:2021iwc} and compared the cutflows. This provided an amelioration of the agreement for intermediate steps, and we observed that the pile-up events were very efficiently removed by the point of applying the two-lepton cut. However, due to the first issue above and the fact that we do not have any efficiency information for the reconstruction of displaced vertices coming from hadronic decays, we could not entirely reproduce each individual step of the cutflows. In the end, our final per-event efficiencies agree nevertheless very well with the ones provided by the ATLAS Collaboration (both with and without pile-up). The complete set of our comparisons of cutflows is given in~\ref{app:cutflows_ALTAS-SUSY-2017-04}.

\subsubsection{Overall selection efficiencies and exclusion limits}

\begin{figure*}\centering
  \includegraphics[width=0.49\textwidth]{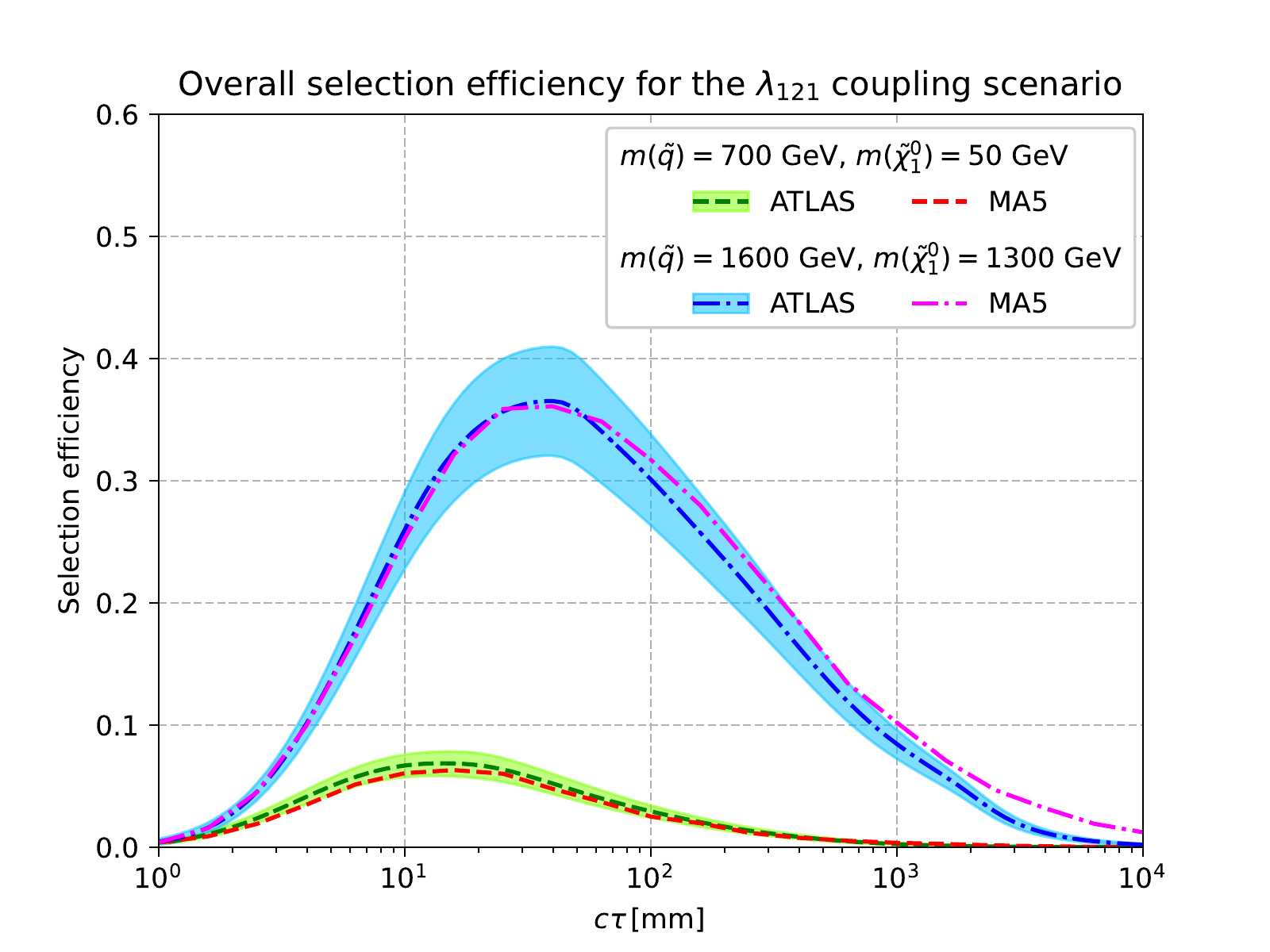}
  \includegraphics[width=0.49\textwidth]{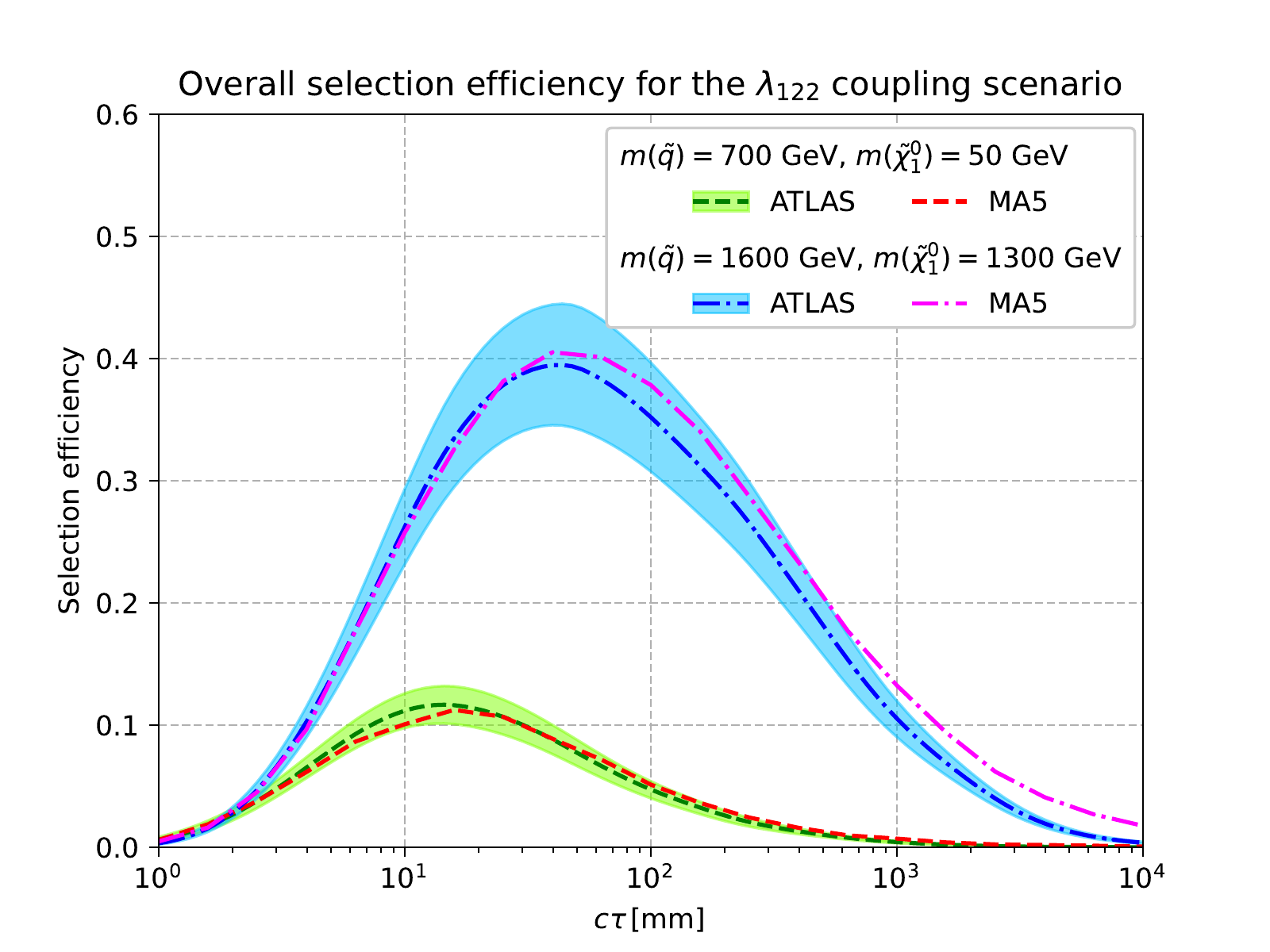} \\
  \includegraphics[width=0.49\textwidth]{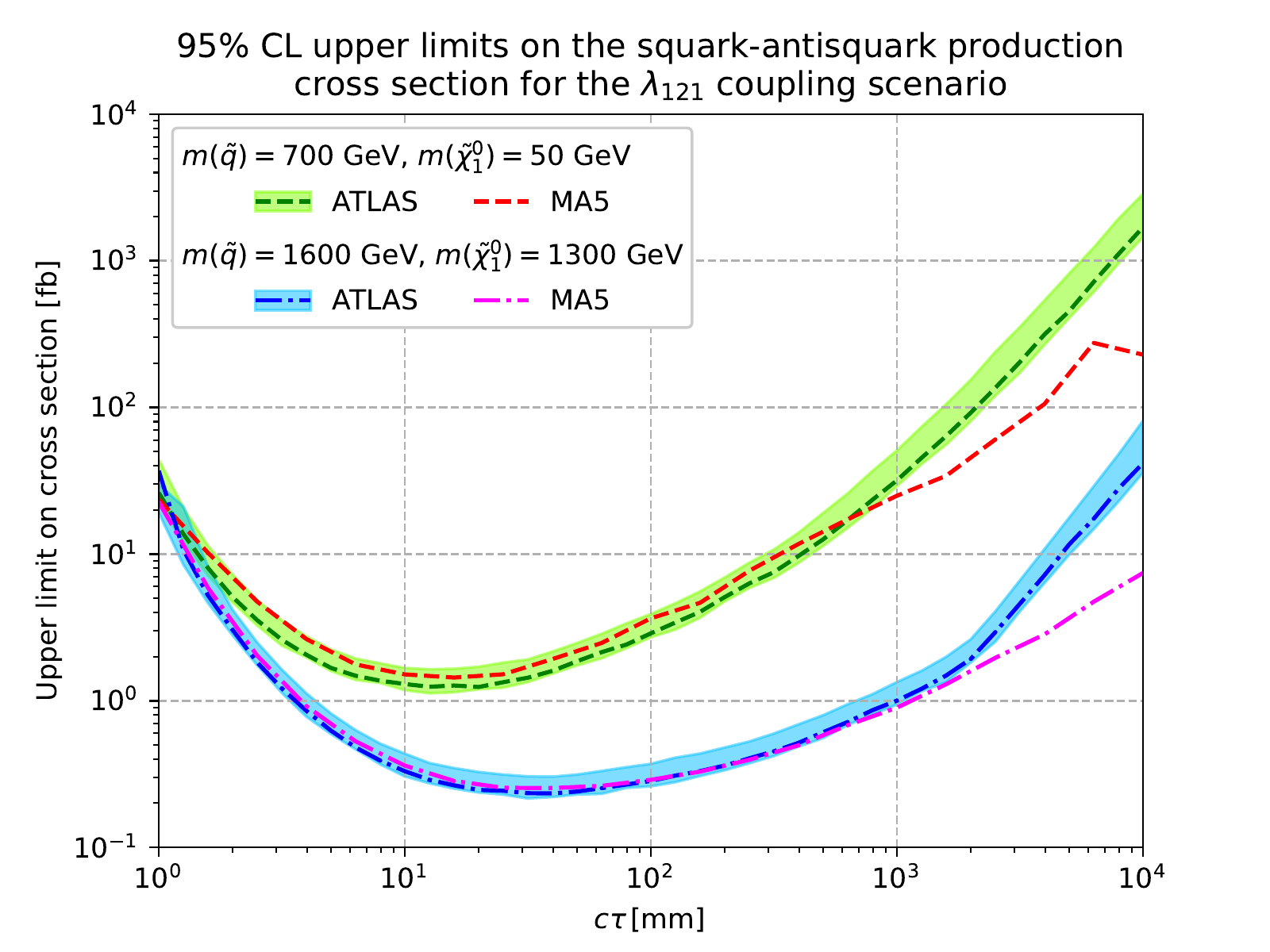}
  \includegraphics[width=0.49\textwidth]{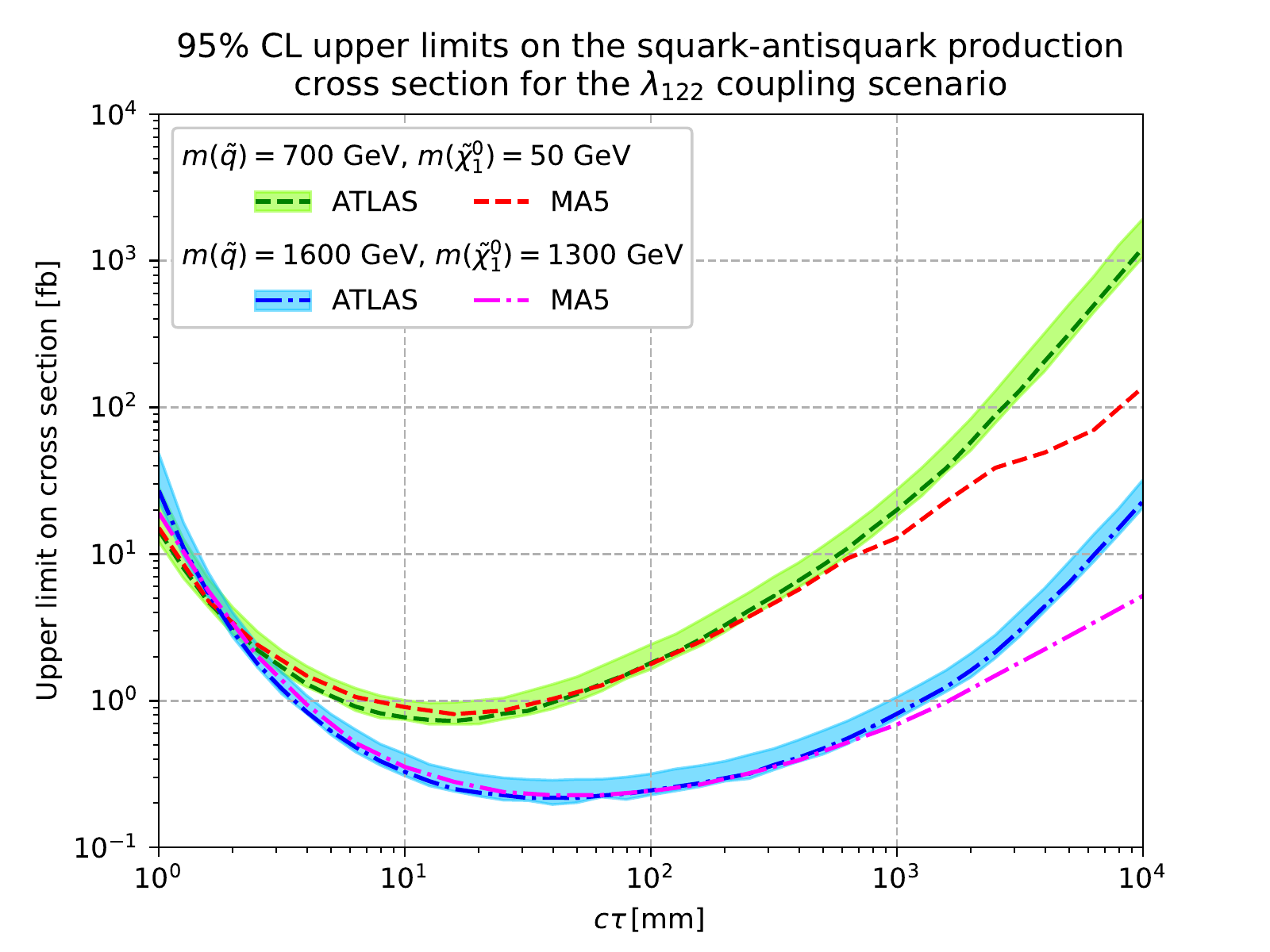}
  \caption{Overall selection efficiencies (top) and upper limits on the squark-antisquark production cross section (bottom) obtained with our {\madanalysis} implementation (MA5), in comparison with the limits (with uncertainties) found by the ATLAS Collaboration. We consider two configurations of squark and neutralino masses in the $\lambda_{121}$ (left) and $\lambda_{122}$ (right) coupling scenario. \label{fig_RPV_scan_plots}}
\end{figure*}

To further validate our implementation, we compared our findings for the lifetime dependence of the selection efficiency, and the upper limits on the squark pair-production cross section for different choices of squark and neutralino masses. We considered $\lambda_{121}$ and $\lambda_{122}$ couplings separately, in accordance with the information available in the \hepdata entry of the analysis. As already mentioned, we choose two configurations of squark and neutralino masses and scan over the lifetime value. Results for the overall selection efficiencies and cross section upper limits are shown in figure~\ref{fig_RPV_scan_plots}.

We can observe that the \madanalysis and ATLAS results are in relatively good agreement up to proper decay lengths of \SI{1000}{mm}. Beyond this value the curves differ. For the $m(\tilde{q})=\SI{700}{GeV}, $ $m(\chi_1^0)=\SI{50}{GeV}$ case at very large lifetimes (above \SI{5000}{mm}) there are kinks in the \madanalysis curves because of very low signal efficiency/insufficient simulated events. We did not simulate more events to improve them, however, for a much more serious reason: the ATLAS curves for $m(\tilde{q})=\SI{1600}{GeV}$ and $m(\chi_1^0)=\SI{1300}{GeV}$ (blue) contain a noticeable kink above $c\tau=\SI{1000}{mm}$, which is particularly striking in the case of the $\lambda_{121}$ scenario. No reasonable explanation could be found in the selection criteria of the analysis, and it is worth mentioning that according to the analysis note, the ATLAS Collaboration did not generate samples for proper decay lengths above \SI{1000}{mm}, but combined different samples in the $c\tau$-range between \SI{10}{mm} and \SI{1000}{mm}. To use these samples for different LLP lifetimes, they reweighted events to account for the different probability of the same events occurring in a sample associated with a different lifetime. For the validation of the {\madanalysis} implementation of the analysis, a different sample was generated for each value of $c\tau$. However, there should be no problem with the reweighting procedure (provided sufficient events are generated). We checked that this also worked for us. 

The analysis fiducial volume only extends to \SI{300}{mm} from the beampipe. For lifetimes above \SI{1000}{mm}, the decays of a heavy particle within the fiducial volume should be roughly uniformly distributed as a function of decay length from the interaction point (with efficiencies that decrease at larger radius), and most LLPs decay outside of it, so there should be very few events with two DVs. Increasing the lifetime of the LLP beyond \SI{1000}{mm} should therefore roughly only lead to an overall linear scaling of the signal efficiency inversely proportional to the lifetime. This is what is observed in the per-decay efficiencies, and what we observe in our results (although this is hard to read from the plots). Moreover, when we combine the per-decay efficiencies provided by the ATLAS Collaboration to give a per-event efficiency according to their prescription, the results also agree with our code. Instead, the results in the official plots seem to be dramatically reduced, differing by a factor of $5$ at \SI{10000}{mm}.

We were not able to find an explanation for this phenomenon. Perhaps events with an additional displaced vertex outside the fiducial region (or even a long way out) are being vetoed by the analysis without this being described anywhere. We reported the issue to the ATLAS SUSY conveners, who were so far not able to provide a solution. The analysis has been taken over by a different group who will investigate this matter once the existing analysis code has been integrated into a new analysis framework, with a view to performing a new analysis with more data. Unfortunately, their conclusions were not yet forthcoming at the time of writing this article, and we therefore regard the results of our analysis for lifetimes above \SI{1000}{mm} to be unvalidated.

\subsection{Long-lived vector-like leptons\label{sec_VLL_LLP}}
\subsubsection{Generalities}

To demonstrate the exclusion potential of the \madanalysis im\-ple\-men\-ta\-tion of the ATLAS LLP search discussed in this section, we shall apply it to a SM extension containing a vector-like pair of lepton doublets. The phenomenology of such models at the LHC is described in \eg\ ref.~\cite{Kumar:2015tna}. For the sake of our signature, unlike in that reference, we shall assume that the new doublet pair $L', \overline{L}'$ (in two-component spinor notation, where $L'$ has the same quantum numbers as the left-handed SM lepton doublet $L$) only mixes with the electron. The corresponding Lagrangian reads
\be
-\mathcal{L} \supset m_{L'} L' \overline{L}' + \epsilon H\cdot L' e_R^3 + y_e^{ij} H \cdot L^i e_R^j + {\rm H.c.},
\ee
where $H$ is the Higgs doublet, $e_R^j$ the set of right-handed lepton $SU(2)_L$ singlets and $y_e^{ij}$ the elements of the usual lepton Yukawa matrices, which we take to be diagonal. The above model is a relatively simple extension of the SM with only two additional parameters, namely the VLL mass $M_{\tau'}$ (which is equal for the charged and neutral states at leading order), and the mixing parameter $\epsilon$. We implemented the model in the \sarah package~\cite{Staub:2013tta,Goodsell:2017pdq}, which we used to generate a custom {\textsc{SPheno}} code~\cite{Porod:2011nf} to compute the particle spectrum and the decay tables. Provided the mixing parameter $\epsilon$ is sufficiently small, the new physical leptons $\tau'$ and $\nu'$ are both long-lived and can be produced at colliders via the processes
\begin{equation}
    pp\to\nu'\overline{\nu}'\,,\quad \nu'\tau'^+\,,\quad \overline{\nu}'\tau'^-\,.
\label{eq_vll_prod_without_jets} \end{equation}

The neutral vector-like lepton (VLL) $\nu'$ then decays to an electron and a $W$-boson, where the latter can further promptly decay to an electron or a muon and a neutrino. This decay chain generates a displaced vertex from which two leptons and a neutrino originate, exactly as in the displaced $\chi_1^0$ decay studied in the RPV SUSY model. This justifies the use of the corresponding efficiencies to assess the constraints on the VLL doublet model.

The behaviour of the new charged lepton $\tau'$ is more complicated. At one loop, electroweak symmetry breaking effects split the neutral and charged states. The charged $\tau'$ can consequently decay to a $\nu'$ and an off-shell $W$-boson, exactly as for the wino model considered in section~\ref{sec_CMS_DT}. Whereas the $\tau'$ decay is associated with a disappearing track signature independently of the $\epsilon$ value, the channel $\tau'\to Ze$ can dominate depending on $\epsilon$. It then becomes possible that the $\tau'$ state has a similar lifetime to the $\nu'$ state and yields DVs from which \emph{three} leptons are issued when the $Z$ boson decays leptonically. We shall however ignore such a long-lived $\tau'$ for the following reasons. In the case of the off-shell $W$-decay dominating, the accurate calculation of the decay to pions (and to a lesser extent the treatment of small mass splittings) is not yet automatic in \sarah (otherwise we could consider a combination with the analysis of the next section). Moreover, in the case of the $Ze$ decay dominating, we have no efficiencies for a three-charged-lepton DV and do not know how that would be interpreted in the analysis (even if those should only consist of a small fraction of the signal rate).

\subsubsection{Bounds from ATLAS-SUSY-2017-04}

We simulate samples of 50,000 signal events by using {\madgraph}~\cite{Alwall:2014hca} (v2.8.3.2) together with {\pythia}~\cite{Sjostrand:2014zea} (v8.244), relying on the UFO model files~\cite{Degrande:2011ua} generated by \sarah~\cite{Staub:2012pb}. The processes considered are the ones given in eq.~\eqref{eq_vll_prod_without_jets}, and the associated matrix elements are allowed to include up to two additional hard jets. They are merged according to the MLM prescription~\cite{Mangano:2001xp,Mangano:2006rw,Alwall:2008qv} with a matching scale set to $M_{\tau'}/4$, and convoluted with the NNPDF~2.3~LO~\cite{Ball:2012cx} set of parton distribution functions. We rescaled the resulting signal cross sections with different $K$-factors ($\nicefrac{2}{3}$, $1$ and $\nicefrac{3}{2}$) to parameterise our ignorance of higher-order correction effects.

In order to assess the impact of the ATLAS-SUSY-2017-04 analysis on the model, we first trade the $\epsilon$ mixing parameter for the proper decay length $c\tau$ of the \emph{neutral} VLL $\nu'$. Next, a scan in the $(M_{\tau'}, c\tau)$ plane is performed, and we evaluate the constraints resulting on each point from our implementation by means of the CL$_s$ method~\cite{Read:2002hq}. We perform a grid scan with mass values below \SI{1600}{GeV} in steps of \SI{50}{GeV}, and proper decay lengths $c\tau$ ranging from 1 to \SI{10000}{mm} with equal spacing of 0.5 on a logarithmic scale (\ie\ powers of 10 increased in steps of 0.5).

\begin{figure}\centering
  \includegraphics[width=0.52\textwidth]{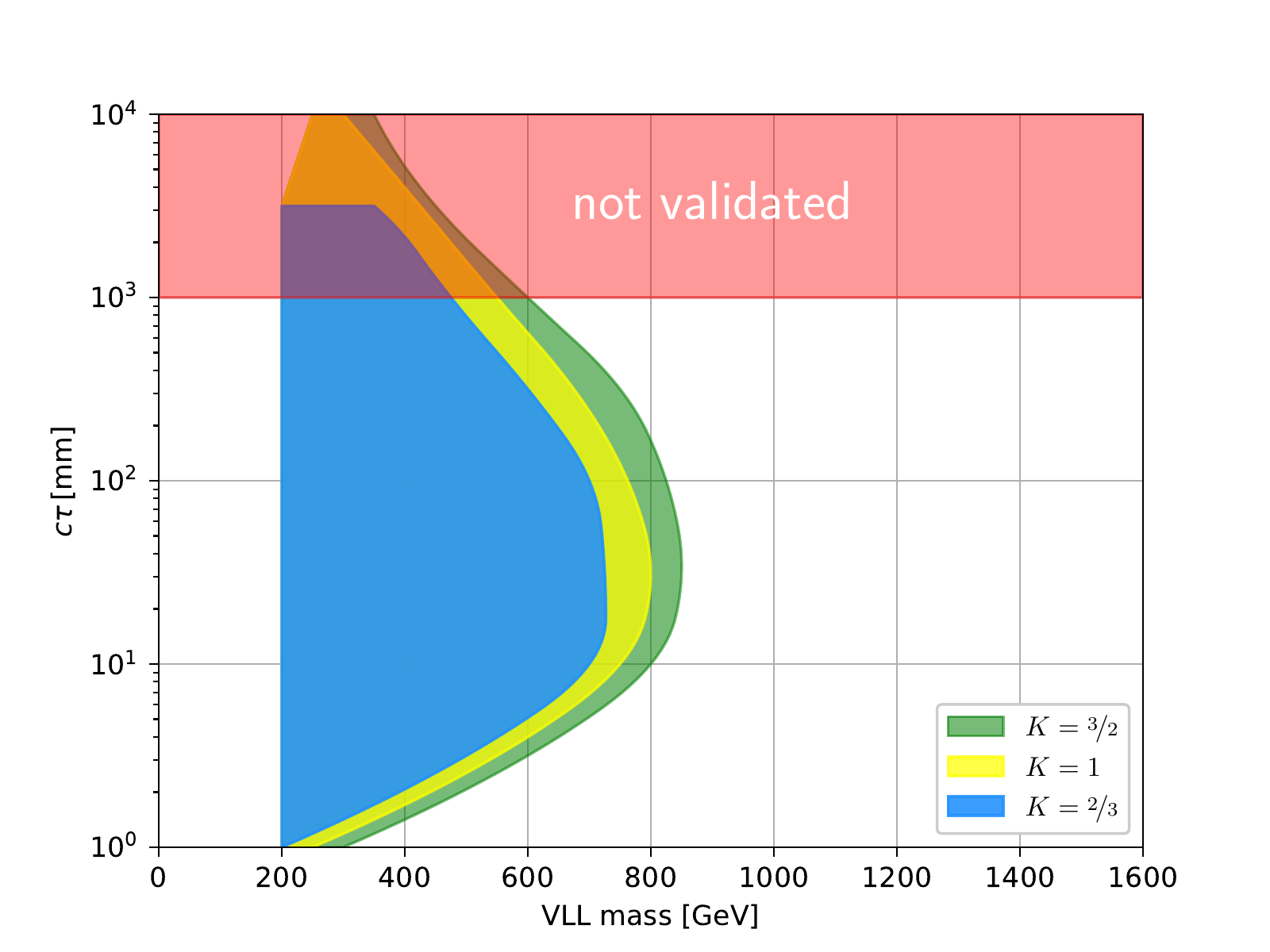}
  \caption{Regions of the VLL model that are excluded at \SI{95}{\percent} confidence level. The results are shown in the $(M_{\tau'}, c\tau)$ plane, and consider different $K$-factors to be applied to the signal rate. The red-shaded region covers decay lengths above $c\tau=\SI{1000}{mm}$, in which there are doubts about the validity of the implementation. \label{VLL_scan_regions}}
\end{figure}

The results are given in figure~\ref{VLL_scan_regions}, the region in which $c\tau>\SI{1000}{mm}$ being highlighted to signal that it is associated with predictions that cannot be trusted due to unsatisfactory validation (see section~\ref{subsec_ATLAS_validation}). No constraints are found for $\nu'$ masses below \SI{200}{GeV}. Moreover, we observe that a neutral VLL $\nu'$ with a decay length $c\tau$ below approximately \SI{2}{mm} is only relatively weakly constrained, its mass being enforced to be above roughly 300 GeV. On the other hand, for higher lifetimes, this lower bound grows to above \SI{700}{GeV}, comparable to the best constraints on long-lived weakly coupled particles. 

\section{Disappearing tracks (CMS-EXO-19-010)\label{sec_CMS_DT}}
\setcounter{equation}{0}
\subsection{Generalities}
Heavy long-lived \emph{charged} particles produced at the LHC show up in the tracking system of the detectors, and then produce a track that ``disappears'' if they decay into heavy invisible states. This classic long-lived particle signature is well motivated by models including heavy electroweak multiplets with component states having masses that are only split after electroweak symmetry breaking, and where the lightest state is the neutral one. Canonical examples include supersymmetric winos and higgsinos~\cite{Chen:1996ap,Ibe:2012hu,Arvanitaki:2012ps,Hall:2012zp,Bobrovskyi:2012dc,Rolbiecki:2015gsa,Goodsell:2021iwc}, and Minimal Dark Matter~\cite{Cirelli:2005uq,Cirelli:2014dsa}.

The ATLAS Collaboration published a search~\cite{ATLAS:2017oal} for new physics when it is manifest through this signature based on an integrated luminosity of $36.1$ \invfb, and they provided together with the analysis results substantial recasting material. The latter includes a pseudocode, and efficiencies for ``strong'' and ``electroweak'' LLP production scenarios that convolve great deal of information about tracklet reconstruction and selection cuts. This formed the basis of a code published in \cite{Belyaev:2020wok} and available on the {\tt LLP Recasting Repository}\cite{llprecasting}, and has also been incorporated into \checkmate~\cite{Desai:2021jsa}. Recently a conference note \cite{ATLAS:2021ttq} analysing the full run 2 dataset of $136$ $\invfb$ appeared, followed by the full paper \cite{ATLAS:2022rme}, which was released while this paper was under review. It would be very interesting in the future to recast that search and compare to the one presented here.

On the other hand, the final CMS analysis of the entire LHC run 2 dataset was already extant~\cite{CMS:2018rea,CMS:2020atg} in early 2020, where the first analysis~\cite{CMS:2018rea} comprised 38.4 \invfb of data from 2015 and 2016, and the second one~\cite{CMS:2020atg} added 101 \invfb from 2017 and 2018. The recasting material consists of cutflows and acceptances for a wide range of LLP masses and lifetimes, but no efficiencies. Moreover, it is a particularly challenging search to recast because the signal regions and many of the cuts are defined in terms of the number of tracker layers that are hit, so the results are dependent on knowledge of the tracker geometry and a method of reproducing the track reconstruction efficiency. This was however achieved in \cite{Goodsell:2021iwc} and released as a public code.\footnote{See the webpages \url{https://github.com/llprecasting/recastingCodes/tree/master/DisappearingTracks/CMS-EXO-19-010} and \url{https://goodsell.pages.in2p3.fr/hackanalysis}.}

We present in the rest of this section the implementation of the code from \cite{Goodsell:2021iwc} in the SFS framework of \madanalysis.

\subsection{Technical details about the SFS implementation}
We refer to ref.~\cite{Goodsell:2021iwc} for the details of the approach taken, the definition of the signal regions and the cuts. We only highlight in this subsection differences that are inherent to the SFS approach.

First, we make use of the new {\tt RecTrackFormat} objects available in the SFS framework, which can comprise the signal as well as charged hadronic particles.

Second, we stress that the analysis uses two specific objects related to the missing transverse energy (MET), namely the classic $p_T^{\rm miss}$ definition and a new variable denoted by $p_T^{\rm miss, \slashed{\mu}}$. The latter corresponds to the missing transverse momentum obtained without including muons in the calculation. This originates from the presence of metastable charged particles that reach the muon system and that are counted as muons for the MET computation. To avoid problems with this, both MET variables are used. However, the MET calculation in the SFS framework only makes use of final-state particles without reference to the detector geometry. In order to accommodate such long-lived particles, we must correct $p_T^{\rm miss}$ by including long-lived track momenta, and we therefore do not need to subtract them from $p_T^{\rm miss, \slashed{\mu}}.$

Finally, due to different parts of the detector that were not functioning and changes in the triggers/cuts over time, the results are split into six data-taking periods, which both in \cite{Goodsell:2021iwc} and \madanalysis require different cutflows. For each of these there are three signal regions (SR1, SR2 and SR3) that differ by the requirements on the number of tracker layers hit, with the exception of the 2015/2016 period, for which only one region (SR3) is relevant. We thus have 12 resulting cutflows: three cutflows for each of the periods 2017, 2018A, 2018B (SR1, SR2 and SR3) and one cutflow for the periods 2015, 2016A, 2016B (SR3). In the \madanalysis implementation, since the total integrated luminosity is fixed for the whole analysis, each cutflow is reweighted by the relative integrated luminosity of the data-taking period as a final cut.

For a specific signal region, the different data-taking periods should be independent. The associated results can therefore be combined with zero correlations. Then in computing the statistics, we have added the possibility of having several sets of separately combined signal regions (so SR1 combined across the 2017, 2017A and 2018B periods, SR2 combined across the same periods and SR3 combined across all six periods) rather than just one combined region per analysis.

\subsection{Validation}
\begin{figure}\centering
  \includegraphics[width=0.49\textwidth]{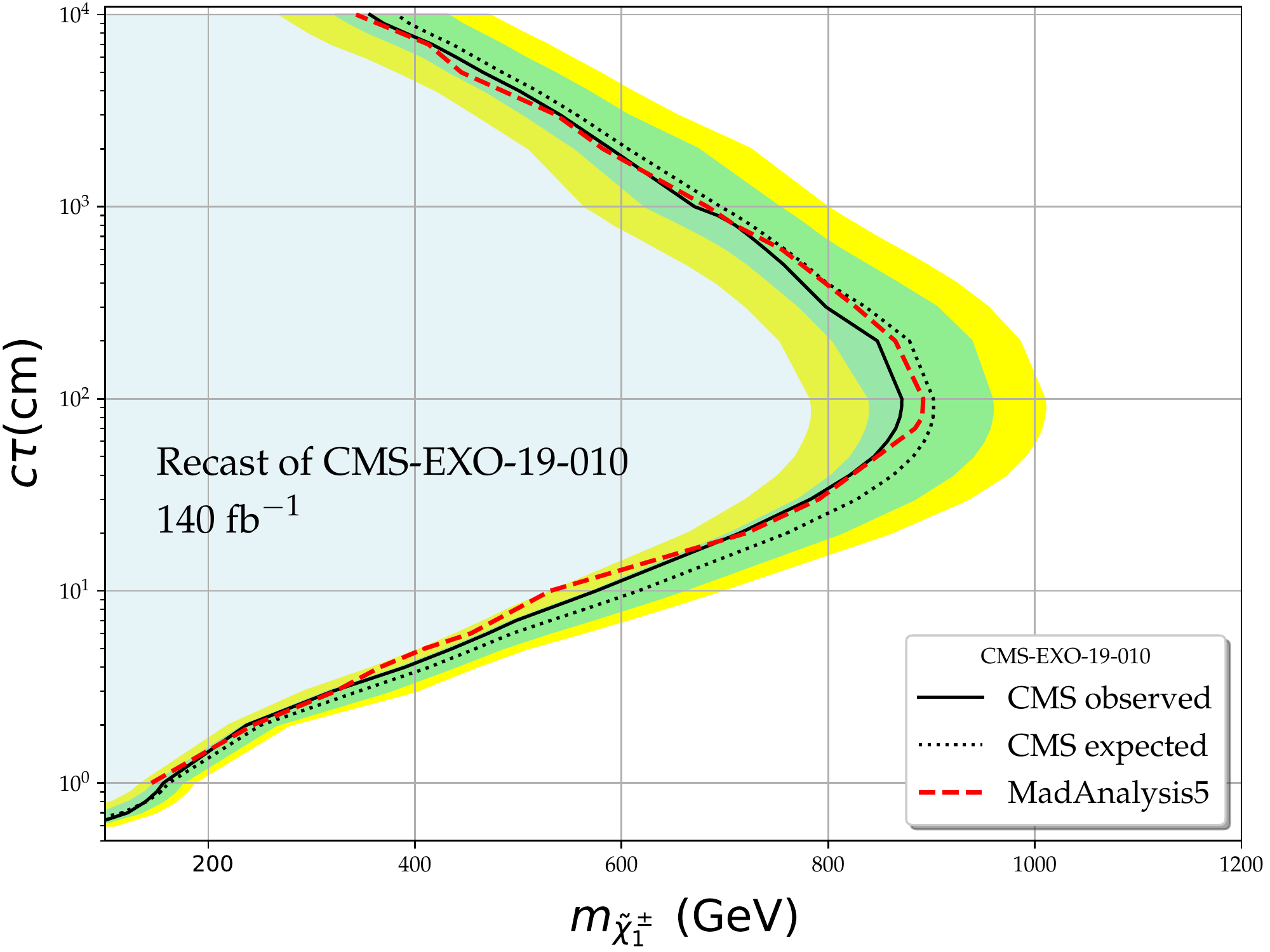}
  \caption{Comparison of the exclusion curve obtained using the {\madanalysis} (SFS) implementation of the CMS disappearing track search and the official curves available from the analysis publication itself. \label{CMS_DT_plot}}
\end{figure}

Ref.~\cite{Goodsell:2021iwc} presented cutflows and various sets of exclusion curves comparing the impact of different details in the simulation chain for the signal (in particular the multiparton jet merging effects in \pythia). We privately checked that the benchmark cutflows were very similar to those produced by {\sc HackAnalysis} and do not reproduce them here. We instead present a single comparison of the exclusion contours obtained with our SFS implementation, which is publicly available on \cite{P82DKS_2021}, to that provided by the CMS Collaboration in their analysis publication~\cite{CMS:2020atg}.

We simulate the production of a pair of long-lived charged winos with \madgraph~\cite{Alwall:2014hca}, making use of the built-in implementation of the MSSM~\cite{Duhr:2011se} to generate the corresponding matrix elements including up to two extra partons. These are then matched with parton showering and hadronisation as performed in \pythia~\cite{Sjostrand:2014zea} using the MLM prescription~\cite{Mangano:2001xp,Mangano:2006rw,Alwall:2008qv} as interfaced in \madgraph. We normalise our signal cross section at NLO+NLL~\cite{Debove:2010kf,Fuks:2012qx,Fuks:2013vua},\footnote{Numerical values are available online from \url{https://twiki.cern.ch/twiki/bin/view/LHCPhysics/SUSYCrossSections}.} and show our results in figure \ref{CMS_DT_plot}. The agreement of the \madanalysis exclusion with the CMS one is even more striking here than in \cite{Goodsell:2021iwc}, perhaps because of the slightly different jet matching or the improved statistical treatment.

\section{Conclusions}\label{sec:conclusion}
We have discussed three classes of new developments of the SFS framework, the simplified fast detector simulator shipped with \madanalysis. First, new options have been added to treat lepton, photon and track isolation, that can now be enforced on the basis of the activity inside a cone centred on the object considered. Second, energy scaling, and in particular jet energy scaling, can be implemented in the same way as any other detector effect that is included in the SFS framework. Finally, we have augmented the code with a particle propagator module that simulates the impact of the magnetic field lying inside the detector. The trajectories of all charged objects and their decay products are hence modified appropriately, in contrast with the previous implementation where they were assumed to propagate along straight lines. To complement this feature that could be important in the treatment of long-lived particles, typical LLP observables like the transverse and longitudinal impact parameters and the coordinates of the point of closest approach are now available within the code, in both its normal mode and expert mode of running.

As examples of usage of these developments, we have implemented in the \madanalysis SFS framework three existing run~2 analyses, namely a CMS run~2 search for displaced leptons in the $e\mu$ channel (CMS-EXO-16-022), the full run~2 CMS search for disappearing tracks (CMS-EXO-19-010) and the partial run~2 ATLAS search for displaced vertices from which a pair of oppositely charged leptons originate (ATLAS-SUSY-2017-04). Our implementations have been carefully validated against public information provided by the ATLAS and CMS Collaborations. While previous attempts have been made to recast the two CMS analyses, the validation here shows the best agreement with the experiments. Moreover, there was no previous recast of the ATLAS search. The codes, available online from the \madanalysis Public Analysis Database\footnote{See the webpage \url{http://madanalysis.irmp.ucl.ac.be/wiki/PublicAnalysisDatabase}.} and as entries in the \madanalysis dataverse~\cite{SVILPJ_2021,31JVGJ_2021,P82DKS_2021}, can hence now be safely used to reinterpret the results of these analyses in the context of any model of physics beyond the SM. As an illustration, we have considered a BSM model in which the SM is extended by a long-lived vector-like lepton doublet, and we have extracted the constraints on its parameter space by recasting the results of the ATLAS-SUSY-2017-04 analysis.

\section*{Acknowledgements}
We thank Brian Francis for very helpful discussions about the CMS disappearing track analysis, and Lakshmi Priya for collaboration on its reinterpretation in \cite{Goodsell:2021iwc}. We thank Sabine Kraml, Humberto Reyes Gonzalez and Sophie Williamson for collaboration on related topics; Andre Lessa for helpful discussions. We thank the organisers of the LLP workshops \url{https://longlivedparticles.web.cern.ch/}. We thank Juliette Alimena for correspondence about \cite{CMS:2021kdm}. 
MDG acknowledges support from the grant
\mbox{``HiggsAutomator''} of the Agence Nationale de la Recherche
(ANR) (ANR-15-CE31-0002). 

\appendix
\section{Impact parameters in the presence of a magnetic field}\label{app:propagation}
In this section, we provide details about the derivation of the impact parameters and the coordinates of the point of closest approach. The resulting expressions are those internally used in the SFS framework, and can be retrieved at the analysis level as described in \ref{app:propexpert}. 

The trajectory of a particle of electric charge $q$, four-momentum $p^\mu = (E/c, {\bf p}(t)) = (E/c, p_x(t), p_y(t), p_z(t))$, and which is subjected to a constant magnetic field $\mathbf{B}$, can be derived from the Lo\-rentz force originating from the field. Within the inner detector volume, we assume a homogeneous magnetic field parallel to the $z$-axis, so we can set ${\bf B} = B {\bf e}_z$. It is then possible to split the description of the transverse motion from the one of the longitudinal motion,
\begin{equation}
  \frac{{\rm d}{\bf p}_T(t)}{{\rm d}t} = \frac{q c^2 B}{E} {\bf p}_T(t)\times{\bf e}_z, \qquad
  \frac{{\rm d}p_z(t)}{{\rm d}t} = 0.
\end{equation}
This shows that the longitudinal momentum $p_z$ and the norm of the transverse momentum $p_T$ are constants of motion, and that the ${\bf p}_T$ two-vector has a circular trajectory in the transverse plane. We get
\begin{equation}\renewcommand{\arraystretch}{1.4}\bsp
  p_x(t) = &\ p_x(t_v)\cos\omega t + p_y(t_v)\sin\omega t,\\
  p_y(t) = &\ p_y(t_v)\cos\omega t - p_x(t_v)\sin\omega t,\\
  p_z(t) = &\ p_z(t_v).
\esp\label{eq:p}\end{equation}
In these expressions, $t_v$ is the moment at which the particle has been created at a displaced vertex located at ${\bf x}_v = (x_v, y_v, z_v)$, and $\omega=q c^2 B/E$ is the cyclotron frequency. Integrating these equations over time leads to
\begin{equation} \renewcommand{\arraystretch}{1.4}\begin{split}
  x(t) =&\ x_h + \frac{R_{\cal H}}{p_T} \big[ p_x(t_v)\sin\omega t-p_y(t_v)\cos\omega t\big],\\
  y(t) =&\ y_h + \frac{R_{\cal H}}{p_T} \big[ p_x(t_v)\cos\omega t+p_y(t_v)\sin\omega t\big],\\
  z(t) =&\ z_v + \frac{c^2 p_z}{E} t.
\end{split}\label{eq:trajectory}\end{equation}
The resulting trajectory is thus a helix ${\cal H}$ of radius $R_{\cal H} = p_T/(q B)$, aligned in the $z$ direction. The coordinates $(x_h, y_h)$ of the centre of the helix in the transverse plane are given by
\begin{equation}
  x_h=x_v+\frac{p_y(t_v)}{qB},
  \qquad\qquad
  y_h=y_v-\frac{p_x(t_v)}{qB}.
\end{equation}
Whereas we should in principle account for synchrotron radiation effects that reduce the kinetic energy of the particle with time, such a phenomenon is neglected.

The position of the point of closest approach stems from the minimisation of the magnitude of the transverse position $x_T(t) = || {\bf x}_T(t)||$ over time,
\begin{equation}
  \left.\frac{{\rm d} x_T(t)}{{\rm d} t}\right|_{t=t_d} = 0.
\end{equation}
We obtain
\begin{equation} \begin{split}
    x_d =&\  x_h \left( 1 - \frac{|R_{\cal H}|}{r_h} \right),\\
    y_d =&\  y_h \left( 1 - \frac{|R_{\cal H}|}{r_h} \right),\\
    z_d =&\  z_v+\frac{p_z}{qB} \arctan (\omega t_d),
\end{split}\label{eq:closestapproach}\end{equation}
where $r_h=\sqrt{x_h^2+y_h^2}$ is the distance of the helix axis from the $z$-axis. Moreover, we have introduced
\begin{equation}\begin{split}
  &\arctan(\omega t_d) \equiv {\tt atan2}\Big( \xi [p_x(t_v)x_h+p_y(t_v)y_h], \\ & \quad 
    \xi [p_x(t_v)y_h-p_y(t_v)x_h] \Big),
\end{split}\label{eq:omegatd}\end{equation}
with $\xi = -{\rm sign}(R_{\cal H})$ and with the arc tangent being expressed through the two-argument function ${\tt atan2}(y,x)$. The latter is equivalent to the argument of a complex number $\arg(x+iy)$, and thus returns values from $-\pi$ to $\pi$. In this way, information about the quadrant in which the complex number $z=x+iy$ is located is kept, and we ensure that the trajectory reaches a minimum and not a maximum of the distance to the $z$-axis at the point of closest approach.

As a consequence of the trajectory being a helix with an axis parallel to the $z$-axis, there are in principle infinitely many points for which $x_T$ is minimised, all those points being equally spaced by a distance $\Delta z = 2\pi p_z(t_v) / (q B)$ on the $z$-axis. We have however chosen the time $t_d$ such that the point of closest approach corresponds to the point that is the closest to the production vertex, as highly energetic particles as produced at hadron colliders most often have only one candidate point for the closest approach lying within the detector volume. We obtain
\begin{equation}
   d_0 = \xi \big(|R_{\cal H}|-r_h\big),\qquad
   d_z = z_v + \frac{p_z}{qB} \arctan(\omega t_d).
\label{eq:impactparameters}\end{equation}
The sign of $d_0$ is derived by rewriting $|d_0|$ as a function of the momentum at the point of closest approach, or more precisely as a function of the $z$-component of the angular momentum $L_z$ at this point ($d_0 = L_z/p_T$).

\section{LLP methods available in the expert mode of \madanalysis}
\label{app:propexpert}
\begin{table*}
  \renewcommand{\arraystretch}{1.4}
  \setlength\tabcolsep{5pt}
  \begin{tabular}{cc|p{8.8cm}}
    Type & Method & Description\\ \hline
    {\tt const MAdouble64\&} & {\tt momentum\_rotation()} & Accessor to the angle of which the momentum of an object has rotated during its helicoidal propagation. This method is only available through the Monte Carlo truth information, obtained from the \texttt{mc()} accessor of the event class.\\
    {\tt const MALorentzVector\&} & {\tt ProductionVertex()} & Accessor to the position four-vector of the vertex $(t_{\rm creation}, {\bf x}_{\rm creation})$ from which the object orignates.\\
    {\tt const MAVector3\&} & {\tt closest\_approach()} & Accessor to the position of the point of closest approach $(x_d, y_d, z_d)$.\\
    {\tt const MAdouble64\&} & {\tt d0()} & Accessor to the transverse impact parameter $d_0$.\\
    {\tt const MAdouble64\&} & {\tt dz()} & Accessor to the longitudinal impact parameter $d_z$.\\
    {\tt const MAdouble64\&} & {\tt d0\_approx()} & Accessor to the transverse impact parameter $\tilde d_0$ as calculated in the absence of a magnetic field.\\
    {\tt const MAdouble64\&} & {\tt dz\_approx()} & Accessor to the longitudinal impact parameter $\tilde d_z$ as calculated in the absence of a magnetic field.\\
    {\tt const vector<IsolationConeType>\&} & {\tt isolCones()} & Accessor to the set of isolation cone objects associated with any instance of a reconstructed lepton, photon and track. For a specific isolation cone object, the number of tracks inside the cone, the sum of their transverse momenta and the sum of the transverse energy deposits inside the cone can be retrieved through the {\tt ntracks()}, {\tt sumPT()} and {\tt sumET()} methods of the {\tt IsolationConeType} class.\\
  \end{tabular}\vspace*{.2cm}
  \caption{List of accessors relevant for LLP analysis implementations in \madanalysis. We indicate the type, in the \madanalysis language, of each method that is given with a brief description. These accessors are available for {\tt RecLeptonFormat}, {\tt RecPhotonFormat} and {\tt RecTrackFormat} objects, with the exception of {\tt momentum\_rotation()} which is a property of the Monte Carlo truth objects.\label{tab:ma5_methods}}
\end{table*}

The internal data format of the \sampleanalyzer core of \madanalysis has been upgraded to facilitate the handling of long-lived objects present in event records. All particle classes are now equipped with methods useful to access properties relevant for LLPs. They are all collected in table~\ref{tab:ma5_methods}.

The accessor {\tt ProductionVertex()} allows users to obtain the position four-vector $(c t_{\rm creation},\ {\bf x}_{\rm creation})$ associated with the vertex at which the object has been produced. Information related to the point of closest approach is available for all Monte Carlo objects of the event through a set of five accessors.

\begin{itemize}
  \item[$\bullet$] The method {\tt closest\_approach()} returns the value of the point of closest approach, \ie\ $(x_d, y_d, z_d)$ in the notation of section~\ref{sec:ma5general}.
  \item[$\bullet$] The method {\tt d0()} returns the value of the transverse impact parameter $d_0$.
  \item[$\bullet$] The method {\tt dz()} returns the value of the longitudinal impact parameter $d_z$.
  \item[$\bullet$] The method {\tt d0\_approx()} returns the value of the approximate transverse impact parameter $\tilde d_0$ that is obtained when neglecting the impact of the magnetic field.
  \item[$\bullet$] The method {\tt dz\_approx()} returns the value of the approximate longitudinal impact parameter $\tilde d_z$ that is obtained when neglecting the impact of the magnetic field.
\end{itemize}

Due to the helicoidal propagation of a specific object, the momentum of the daughter particles originating from the object's decay will rotate as well. The corresponding rotation angle $\varphi$ can be obtained through the {\tt momentum\_rota\-ti\-on()} method of the object's Monte Carlo information. The latter is accessible through the \texttt{mc()} accessor and is filled on runtime when SFS detector simulation is turned on.

In addition, we have upgraded the SFS module to ameliorate the implementation of object isolation at the analysis level. The SFS has cannibalised the {\tt Isolation\-ConeType} class and the associated {\tt isolCone()} method that were available, but deprecated, for certain reconstructed objects.\footnote{The {\tt Isolation\-ConeType} class was initially developed as a patch to missing functionalities in \delphes more than 5 years ago. This was part of what was called the `MA5tune' of \delphes in ref.~\cite{Dumont:2014tja}. These functionalities have however been added to \delphes in the meantime, making the MA5tune of \delphes deprecated. For backwards compatibility, they have nevertheless been kept in \sampleanalyzer.} Thanks to this, users can again determine whether photons, leptons and tracks are isolated through the {\tt isolCones()} accessor of the corresponding {\tt RecLeptonFormat}, {\tt RecPhotonFormat} and {\tt RecTrack\-Format} objects. This returns a vector of {\tt I\-so\-la\-tion\-ConeType} objects. Each of the components of this vector is an {\tt I\-so\-la\-tion\-ConeType} object that includes, for a given cone radius $\Delta R$ (to be fixed when defining the detector), the sum of the transverse momenta of the objects inside the cone ({\tt sumPT()}), as well as the total transverse energy inside the cone ({\tt sumET()}).

\section{Validation of the ATLAS-SUSY-2017-04 SFS implementation}
\label{app:cutflows_ALTAS-SUSY-2017-04}

\setlength{\tabcolsep}{8pt}
\renewcommand{\arraystretch}{1.2}
\begin{table*}\centering
  \begin{tabular}{c||rr|rr|rr||rr|rr|rr}
    &\multicolumn{6}{c||}{\textbf{\boldmath $c\tau=\SI[detect-weight]{30}{mm}$}}&\multicolumn{6}{c}{\textbf{\boldmath $c\tau=\SI[detect-weight]{1000}{mm}$}}\\
    &\multicolumn{6}{c||}{$N_{\text{weighted}}$ (SFS/{\color{blue}ATLAS})}&\multicolumn{6}{c}{$N_{\text{weighted}}$ (SFS/{\color{blue}ATLAS})}\\
    Channel&\multicolumn{2}{c}{$ee$}&\multicolumn{2}{c}{$e\mu$}&\multicolumn{2}{c||}{$\mu\mu$}&\multicolumn{2}{c}{$ee$}&\multicolumn{2}{c}{$e\mu$}&\multicolumn{2}{c}{$\mu\mu$}\\\hline
    {\small No cuts}&$21.0$&{\color{blue}$21.0$}&$21.2$&{\color{blue}$21.2$}&$20.8$&{\color{blue}$20.8$}&$21.1$&{\color{blue}$21.1$}&$20.9$&{\color{blue}$20.9$}&$21.1$&{\color{blue}$21.1$}\\
    {\small Triggers}&$21.0$&{\color{blue}$20.6$}&$21.0$&{\color{blue}$20.5$}&$20.2$&{\color{blue}$16.6$}&$21.1$&{\color{blue}$19.1$}&$20.9$&{\color{blue}$18.5$}&$21.0$&{\color{blue}$10.9$}\\
    {\small Cosmic-ray veto}&$21.0$&{\color{blue}$20.6$}&$21.0$&{\color{blue}$20.5$}&$20.2$&{\color{blue}$16.6$}&$21.1$&{\color{blue}$19.0$}&$20.8$&{\color{blue}$18.4$}&$21.0$&{\color{blue}$10.9$}\\
    {\color{red}\small Primary vertex}&{\color{red}$21.0$}&{\color{blue}$20.6$}&{\color{red}$21.0$}&{\color{blue}$20.5$}&{\color{red}$20.2$}&{\color{blue}$16.6$}&{\color{red}$21.1$}&{\color{blue}$19.0$}&{\color{red}$20.8$}&{\color{blue}$18.4$}&{\color{red}$21.0$}&{\color{blue}$10.9$}\\
    {\small $N(\text{DV})\geq 1$}&$9.2$&{\color{blue}$15.1$}&$9.5$&{\color{blue}$15.2$}&$9.3$&{\color{blue}$12.7$}&$3.9$&{\color{blue}$10.5$}&$4.1$&{\color{blue}$10.3$}&$4.3$&{\color{blue}$6.6$}\\\hline
    {\small Vertex fit}&$9.2$&{\color{blue}$15.1$}&$9.5$&{\color{blue}$15.2$}&$9.3$&{\color{blue}$12.7$}&$3.9$&{\color{blue}$10.5$}&$4.1$&{\color{blue}$10.2$}&$4.3$&{\color{blue}$6.6$}\\
    {\small $r_{xy}$}&$9.2$&{\color{blue}$15.1$}&$9.5$&{\color{blue}$15.2$}&$9.3$&{\color{blue}$12.7$}&$3.9$&{\color{blue}$10.5$}&$4.1$&{\color{blue}$10.2$}&$4.3$&{\color{blue}$6.6$}\\
    {\small Fiducial volume}&$9.2$&{\color{blue}$14.8$}&$9.5$&{\color{blue}$14.9$}&$9.3$&{\color{blue}$12.4$}&$3.4$&{\color{blue}$9.8$}&$3.6$&{\color{blue}$9.6$}&$3.7$&{\color{blue}$6.3$}\\
    {\small Dis. pixel mod. veto}&$9.0$&{\color{blue}$14.4$}&$9.2$&{\color{blue}$14.6$}&$9.0$&{\color{blue}$12.2$}&$3.2$&{\color{blue}$9.3$}&$3.4$&{\color{blue}$9.1$}&$3.4$&{\color{blue}$5.9$}\\
    {\small Material veto}&$7.9$&{\color{blue}$10.9$}&$8.2$&{\color{blue}$11.1$}&$8.6$&{\color{blue}$12.2$}&$2.3$&{\color{blue}$4.9$}&$2.4$&{\color{blue}$4.8$}&$3.0$&{\color{blue}$5.9$}\\
    {\small $N(l)\geq 1$}&$7.9$&{\color{blue}$7.7$}&$8.2$&{\color{blue}$9.5$}&$8.6$&{\color{blue}$8.5$}&$2.3$&{\color{blue}$2.2$}&$2.4$&{\color{blue}$2.6$}&$3.0$&{\color{blue}$2.7$}\\
    {\small $N(l)\geq 2$}&$7.9$&{\color{blue}$5.4$}&$8.2$&{\color{blue}$6.3$}&$8.6$&{\color{blue}$6.8$}&$2.3$&{\color{blue}$1.3$}&$2.4$&{\color{blue}$1.5$}&$3.0$&{\color{blue}$2.0$}\\
    {\color{red}\small Lepton kinematics}&{\color{red}$7.9$}&{\color{blue}$5.3$}&{\color{red}$8.2$}&{\color{blue}$6.3$}&{\color{red}$8.6$}&{\color{blue}$6.8$}&{\color{red}$2.3$}&{\color{blue}$1.2$}&{\color{red}$2.4$}&{\color{blue}$1.5$}&{\color{red}$3.0$}&{\color{blue}$2.0$}\\
    {\color{red}\small Lepton identification}&{\color{red}$7.9$}&{\color{blue}$4.7$}&{\color{red}$8.2$}&{\color{blue}$5.5$}&{\color{red}$8.6$}&{\color{blue}$6.2$}&{\color{red}$2.3$}&{\color{blue}$1.1$}&{\color{red}$2.4$}&{\color{blue}$1.3$}&{\color{red}$3.0$}&{\color{blue}$1.8$}\\
    {\small Overlap removal}&$7.9$&{\color{blue}$4.7$}&$8.2$&{\color{blue}$5.4$}&$8.6$&{\color{blue}$6.2$}&$2.3$&{\color{blue}$1.1$}&$2.4$&{\color{blue}$1.3$}&$3.0$&{\color{blue}$1.8$}\\
    {\small Trigger matching}&$7.7$&{\color{blue}$4.7$}&$8.0$&{\color{blue}$5.2$}&$8.3$&{\color{blue}$5.7$}&$2.2$&{\color{blue}$1.1$}&$2.4$&{\color{blue}$1.2$}&$2.8$&{\color{blue}$1.7$}\\
    {\small Presel. matching}&$7.7$&{\color{blue}$4.7$}&$8.0$&{\color{blue}$5.2$}&$8.3$&{\color{blue}$5.7$}&$2.2$&{\color{blue}$1.1$}&$2.4$&{\color{blue}$1.2$}&$2.8$&{\color{blue}$1.7$}\\
    $m_{\text{DV}}$&$7.7$&{\color{blue}$4.7$}&$8.0$&{\color{blue}$5.2$}&$8.3$&{\color{blue}$5.7$}&$2.2$&{\color{blue}$1.1$}&$2.4$&{\color{blue}$1.2$}&$2.8$&{\color{blue}$1.7$}\\
    {\small Opposite charge}&$4.6$&{\color{blue}$4.6$}&$5.2$&{\color{blue}$5.1$}&$5.7$&{\color{blue}$5.7$}&$1.2$&{\color{blue}$1.0$}&$1.3$&{\color{blue}$1.2$}&$2.0$&{\color{blue}$1.7$}\\
  \end{tabular}
  \caption{Cutflow tables for two RPV SUSY scenarios in which all decays induced by the $\lambda_{121}$ and $\lambda_{122}$ couplings are included with an equal branching ratio. We classify the results according to the flavour of the leptons originating from the displaced vertices, the neutralino proper decay lengths are fixed to \SI{30}{mm} (left) and \SI{1000}{mm} (right), and the squark and neutralino masses are respectively fixed to \SI{1600}{GeV} and \SI{1300}{GeV} in both cases. The numbers $N_{\text{weighted}}$ are reweighted to match to the relevant squark pair production cross section and an integrated luminosity of \SI{32.8}{fb^{-1}}, the initial number of events matching thus the one provided by the ATLAS Collaboration. The cuts highlighted in red are not implemented in the SFS code. \label{tab_cutflow_RPV}}
\end{table*}

\begin{table*}\centering
  \begin{tabular}{c|rr|rr|rr}
    &\multicolumn{6}{c}{\textbf{\boldmath $m(Z')=\SI[detect-weight]{100}{GeV}$}}\\
    &\multicolumn{6}{c}{$N_{\text{weighted}}$ (SFS/{\color{blue}ATLAS})}\\
    Channel&\multicolumn{2}{c}{$ee$}&\multicolumn{2}{c}{$e\mu$}&\multicolumn{2}{c}{$\mu\mu$}\\\hline
    No cuts&$20000.0$&{\color{blue}$20000.0$}&$20000.0$&{\color{blue}$20000.0$}&$20000.0$&{\color{blue}$20000.0$}\\
    Triggers&$301.4$&{\color{blue}$322.7$}&$400.5$&{\color{blue}$462.1$}&$757.9$&{\color{blue}$790.8$}\\
    Cosmic-ray veto&$301.4$&{\color{blue}$322.7$}&$400.5$&{\color{blue}$462.1$}&$757.9$&{\color{blue}$790.8$}\\
    {\color{red}Primary vertex}&$301.4$&{\color{blue}$322.7$}&$400.5$&{\color{blue}$462.1$}&$757.9$&{\color{blue}$790.8$}\\
    $N(\text{DV})\geq 1$&$31.1$&{\color{blue}$124.8$}&$45.9$&{\color{blue}$182.6$}&$124.8$&{\color{blue}$398.1$}\\\hline
    Vertex fit&$31.1$&{\color{blue}$124.8$}&$45.9$&{\color{blue}$182.6$}&$124.8$&{\color{blue}$398.1$}\\
    $r_{xy}$&$31.1$&{\color{blue}$124.8$}&$45.9$&{\color{blue}$182.6$}&$124.8$&{\color{blue}$398.1$}\\
    Fiducial volume&$23.0$&{\color{blue}$121.1$}&$40.9$&{\color{blue}$176.0$}&$116.8$&{\color{blue}$375.7$}\\
    Dis. pixel mod. veto&$23.0$&{\color{blue}$120.3$}&$39.9$&{\color{blue}$164.7$}&$102.8$&{\color{blue}$362.7$}\\
    Material veto&$18.0$&{\color{blue}$71.4$}&$30.9$&{\color{blue}$106.2$}&$102.8$&{\color{blue}$362.7$}\\
    $N(l)\geq 1$&$18.0$&{\color{blue}$44.7$}&$30.9$&{\color{blue}$69.8$}&$102.8$&{\color{blue}$253.4$}\\
    $N(l)\geq 2$&$18.0$&{\color{blue}$38.6$}&$30.9$&{\color{blue}$61.8$}&$102.8$&{\color{blue}$246.7$}\\
    {\color{red}Lepton kinematics}&$18.0$&{\color{blue}$37.7$}&$30.9$&{\color{blue}$60.0$}&$102.8$&{\color{blue}$246.2$}\\
    {\color{red}Lepton identification}&$18.0$&{\color{blue}$35.9$}&$30.9$&{\color{blue}$57.5$}&$102.8$&{\color{blue}$238.2$}\\
    Overlap removal&$18.0$&{\color{blue}$35.9$}&$30.9$&{\color{blue}$57.5$}&$102.8$&{\color{blue}$238.2$}\\
    Trigger matching&$18.0$&{\color{blue}$35.9$}&$30.9$&{\color{blue}$57.5$}&$102.8$&{\color{blue}$237.3$}\\
    Presel. matching&$9.0$&{\color{blue}$14.8$}&$21.0$&{\color{blue}$20.1$}&$81.8$&{\color{blue}$79.5$}\\
    $m_{\text{DV}}$&$9.0$&{\color{blue}$14.8$}&$21.0$&{\color{blue}$20.1$}&$81.8$&{\color{blue}$79.5$}\\
    Opposite charge&$9.0$&{\color{blue}$14.8$}&$21.0$&{\color{blue}$20.1$}&$81.8$&{\color{blue}$79.5$}\\
  \end{tabular}
  \caption{Cutflow tables associated with the $Z'$ toy model, for a $Z'$ mass fixed to \SI{100}{GeV}. We classify the results according to the flavour of the leptons originating from the displaced vertices, and the numbers of events $N_{\text{weighted}}$ correspond to the number of events generated once the initial number of events is fixed to 20,000. The cuts highlighted in red are not used in the SFS code.
    \label{tab_cutflow_ZPrime100GeV}}
\end{table*}

\begin{table*}\centering
  \begin{tabular}{l|rr|rr|rr}
    &\multicolumn{6}{c}{\textbf{\boldmath $m(Z')=\SI[detect-weight]{1000}{GeV}$}}\\
    &\multicolumn{6}{c}{$N_{\text{weighted}}$ (SFS/{\color{blue}ATLAS})}\\
    Channel&\multicolumn{2}{c}{$ee$}&\multicolumn{2}{c}{$e\mu$}&\multicolumn{2}{c}{$\mu\mu$}\\\hline
    No cuts&$20000.0$&{\color{blue}$20000.0$}&$20143.6$&{\color{blue}$20143.6$}&$19608.3$&{\color{blue}$19608.3$}\\
    Triggers&$19029.3$&{\color{blue}$17871.4$}&$18323.0$&{\color{blue}$16465.8$}&$10239.6$&{\color{blue}$9657.6$}\\
    Cosmic-ray veto&$19024.3$&{\color{blue}$17864.4$}&$18314.9$&{\color{blue}$16457.3$}&$10239.6$&{\color{blue}$9655.8$}\\
    {\color{red}Primary vertex}&$19024.3$&{\color{blue}$17858.5$}&$18314.9$&{\color{blue}$16453.6$}&$10239.6$&{\color{blue}$9655.0$}\\
    $N(\text{DV})\geq 1$&$3255.0$&{\color{blue}$6457.8$}&$3964.7$&{\color{blue}$6376.3$}&$3208.0$&{\color{blue}$4199.3$}\\\hline
    Vertex fit&$3255.0$&{\color{blue}$6455.9$}&$3964.7$&{\color{blue}$6373.9$}&$3208.0$&{\color{blue}$4197.6$}\\
    $r_{xy}$&$3255.0$&{\color{blue}$6455.9$}&$3964.7$&{\color{blue}$6373.9$}&$3208.0$&{\color{blue}$4196.7$}\\
    Fiducial volume&$2343.1$&{\color{blue}$5986.3$}&$2849.8$&{\color{blue}$5960.4$}&$2401.1$&{\color{blue}$3969.0$}\\
    Dis. pixel mod. veto&$2319.1$&{\color{blue}$5759.8$}&$2810.5$&{\color{blue}$5791.1$}&$2381.4$&{\color{blue}$3858.0$}\\
    Material veto&$2083.2$&{\color{blue}$3848.9$}&$2514.4$&{\color{blue}$4065.3$}&$2381.4$&{\color{blue}$3858.0$}\\
    $N(l)\geq 1$&$2083.2$&{\color{blue}$2340.4$}&$2514.4$&{\color{blue}$2816.9$}&$2381.4$&{\color{blue}$2453.4$}\\
    $N(l)\geq 2$&$2083.2$&{\color{blue}$2192.9$}&$2514.4$&{\color{blue}$2654.8$}&$2381.4$&{\color{blue}$2342.2$}\\
    {\color{red}Lepton kinematics}&$2083.2$&{\color{blue}$2180.9$}&$2514.4$&{\color{blue}$2645.4$}&$2381.4$&{\color{blue}$2340.9$}\\
    {\color{red}Lepton identification}&$2083.2$&{\color{blue}$2113.7$}&$2514.4$&{\color{blue}$2499.4$}&$2381.4$&{\color{blue}$2215.6$}\\
    Overlap removal&$2083.2$&{\color{blue}$2113.7$}&$2514.4$&{\color{blue}$2497.1$}&$2381.4$&{\color{blue}$2215.6$}\\
    Trigger matching&$2083.2$&{\color{blue}$2113.7$}&$2514.4$&{\color{blue}$2497.1$}&$2381.4$&{\color{blue}$2173.5$}\\
    Presel. matching&$2083.2$&{\color{blue}$2110.8$}&$2501.3$&{\color{blue}$2480.9$}&$2381.4$&{\color{blue}$2170.1$}\\
    $m_{\text{DV}}$&$2083.2$&{\color{blue}$2110.8$}&$2500.3$&{\color{blue}$2480.9$}&$2381.4$&{\color{blue}$2170.1$}\\
    Opposite charge&$2083.2$&{\color{blue}$2088.4$}&$2500.3$&{\color{blue}$2468.4$}&$2381.4$&{\color{blue}$2166.0$}\\\
  \end{tabular}
  \caption{Same as Table~\ref{tab_cutflow_ZPrime100GeV} for a $Z'$ mass of \SI{1000}{GeV}.\label{tab_cutflow_ZPrime1000GeV}}
\end{table*}

Tables~\ref{tab_cutflow_RPV}, \ref{tab_cutflow_ZPrime100GeV} and \ref{tab_cutflow_ZPrime1000GeV} show the number of events surviving the different cuts of the ATLAS-SUSY-2017-04 analysis. The cutflows of the RPV SUSY model are reweighted to represent the expected number of events, given the relevant squark-antisquark production cross section and an integrated luminosity of \SI{32.8}{fb^{-1}}. In the cutflows of the $Z'$ toy model, the numbers represent the number of events generated. In practice, all {\madanalysis} predictions correspond to initial weights before all cuts set to the corresponding ATLAS numbers.

The cuts highlighted in red are not implemented in our SFS implementation, \ie\ they select all events. The first of these cuts is called ``{\it Primary vertex}'' and was discussed in section~\ref{subsec_ATLAS_event_level_requirements}. Monte Carlo truth events containing only simulated signal include a primary vertex at the origin of the coordinate system by definition, so this cut would be redundant; and from the cutflows we see that the efficiency of the primary vertex reconstruction within ATLAS must be very high anyway.  The cuts ``{\it Lepton kinematics}'' and ``{\it Lepton identification}'' are related to lepton reconstruction, which is trivial in the Monte Carlo truth in the absence of specific provided efficiencies. Moreover, the kinematic requirements relevant for the particles associated with displaced vertices are also covered by the preselection.

\bibliography{llp_ma5}
\end{document}